\documentclass[%
 reprint,
showpacs,
 amsmath,amssymb,
 aps,
prb,
]{revtex4-1}

\usepackage{graphicx}
\usepackage{dcolumn}
\usepackage{bm}

\usepackage{amsmath}
\usepackage{amsthm}

\usepackage[english]{babel}
\usepackage[utf8]{inputenc}
\usepackage{graphicx}
\usepackage[colorinlistoftodos]{todonotes}
\usepackage{bbold }
\usepackage{physics}
\usepackage[utf8]{inputenc}
\usepackage{amssymb }
\usepackage{graphicx}
\DeclareGraphicsExtensions{.eps}
\usepackage{epstopdf}
\usepackage{latexsym}
\usepackage{amsfonts}
\usepackage{xcolor}
\usepackage{units}
\usepackage{bm}
\usepackage{verbatim}
\usepackage[colorlinks = true,
            linkcolor = red,
            urlcolor  = blue,
            citecolor = magenta,
            anchorcolor = red]{hyperref}
\usepackage{esint}
\usepackage{soul}
\usepackage{cancel}
\usepackage[normalem]{ulem}
\usepackage{empheq}
\usepackage{dsfont}

\newcommand*{\mybox}[1]{\framebox{#1}}

\definecolor{AB-color}{RGB}{128,0,128}
\begin{document}


\title{Manipulation of Cooper pair entanglement in hybrid topological Josephson junctions}

\author{Gianmichele Blasi}
\email{gianmichele.blasi}
\author{Fabio Taddei}
\author{Vittorio Giovannetti}
\author{Alessandro Braggio}%
\email{alessandro.braggio}
\affiliation{%
NEST, Scuola Normale Superiore and Istituto Nanoscienze-CNR, I-56126 Pisa, Italy}

%
%

\date{\today}

\begin{abstract}
In this work we investigate the supercurrent in a hybrid topological Josephson junction consisting of two planes of topological insulator (TI) in a specific configuration, which allows both local (LAR) and crossed (CAR) Andreev processes at the interfaces with two conventional s-wave superconductors.
We describe the effects of gate voltage and magnetic flux controls applied to the edge states of each TI. In particular, we demonstrate that the voltage gating allows the manipulation of the entaglement symmetry of non-local Cooper pairs associated to the CAR process.
We establish a connection between the Josephson current-phase relationship of the system and the action of the two external fields, finding that they selectively modify the LAR or the CAR contributions.
Remarkably, we find that the critical current of the junction takes a very simple form which reflects the change in the symmetry occurred to the entangled state and allows to determine the microscopic parameters of the junction.
\end{abstract}

\pacs{73.23.-b, 03.67.Bg, 74.45.+c }
\maketitle


\begin{figure}[t]
\includegraphics[width=.5\textwidth]{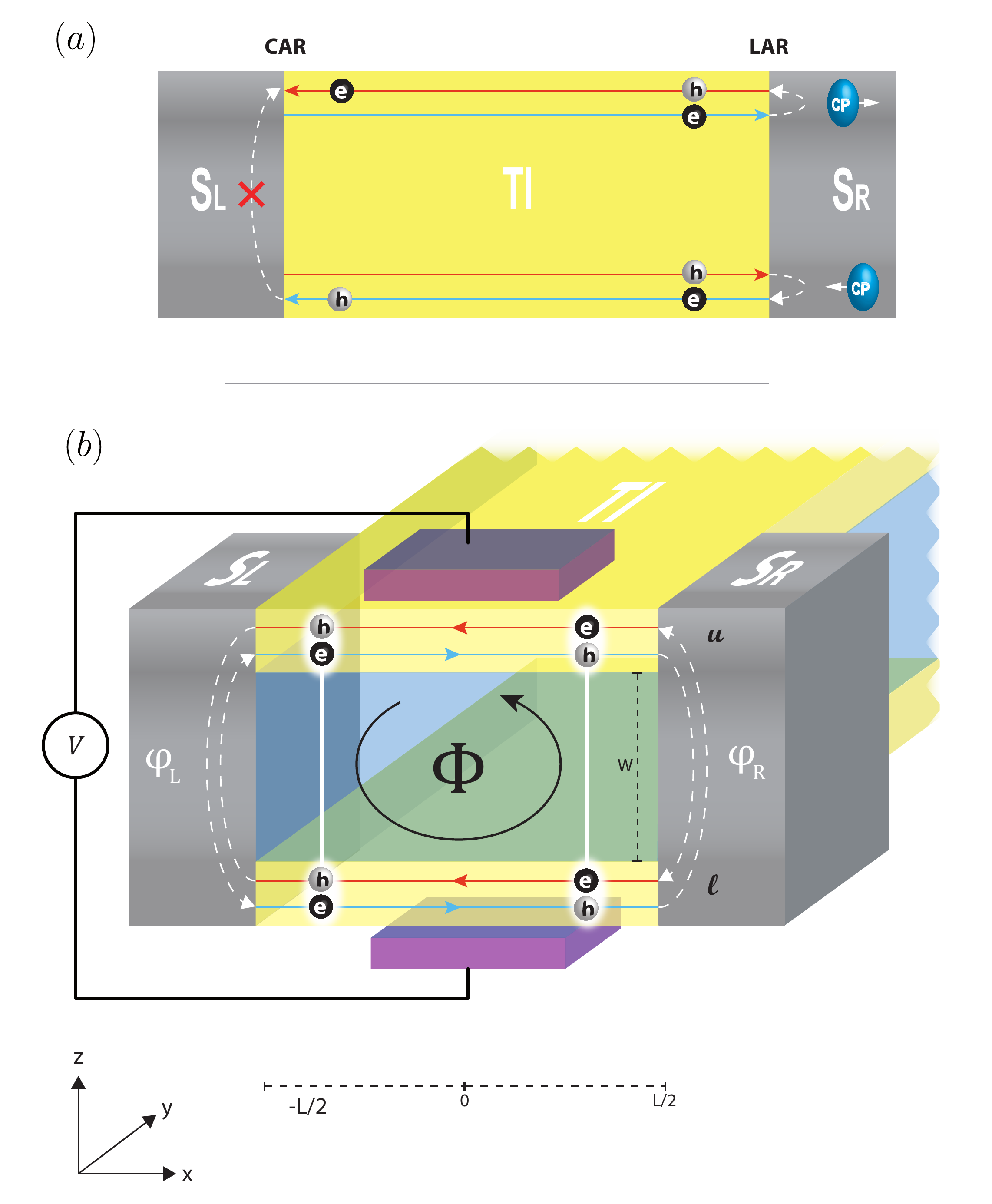}
\centering
\caption{$(a)$ Standard $S$-$TI$-$S$ junction: $S_{L/R}$ superconductors (gray) and the TI (yellow). The arrowed blue (red) solid lines represent the 1D helical edge states with spin $\uparrow$ ($\downarrow$). At interface with $S_R$, two example of emission/absorption of CPs due to LARs. The impossibility of a CAR process where, putatively, $\uparrow$-h in the lower-edge is Andreev reflected in a $\downarrow$-e is represented at at $S_L$ interface with a cross.  $(b)$ Proposed double TI junction - 3D scheme. It consists of an heterostructure (say CdTe-HgTe) grown along the $z$-axis resulting in two layers of TIs. Edge modes running in the backside part of the device here are not represented. On the frontal side, in the $x$-$z$-plane of the scheme, are depicted two CAR processes where CP are non-locally splitted. The application of the $V$- and $\Phi$-field due to the presence of side gates and the induced magnetic flux. Such a fields act in terms of the unitary operators $\mathcal{U}^{\rm u}_{\rm \Phi}(\theta_{\rm \Phi})$ and $\mathcal{U}^{\rm \ell}_{\rm V}(\theta_{\rm V})$ (see text) on the upper and lower edges respectively. Other configurations of local fields can be considered and in Appendix~\ref{Appendix_D} it is indeed shown they are fully equivalent.
} 
    \label{3D} 
\end{figure}

\section{Introduction} Quantum mechanics may revolutionize the way we encode, transmit and elaborate the information. A crucial element is the capability to generate and  manipulate entangled states \cite{Einstein1935,Bell1966,Clauser1969}. First successful steps has been performed on photons \cite{Aspect1981,Tittel1998,Ekert1991,Gisin2002}. 
To deal with quantum technology and the development of a quantum computer, though, one needs to bring those capabilities in the solid state platform to afford the embeddability and scalability issues \cite{Loss1998,Nazarov2005,Burkard2000,Recher2001,Bena2002,Kim2004,hussein2017,hussein2016}.
In Refs.~\onlinecite{Sato2010,Chen2012,Virtanen2012,Reinthaler2013,Choi2014,Sato2014,crepin2014,Veldhorst2014,Zhang2015,Strom2015,Wang2015,Hou2016,Islam2017}  2D topological insulators (TIs) has been put forward for the production and detection of spin-entangled singlet Cooper pairs originating in s-wave superconductors.
2D TIs are materials characterized by edge state modes with helical nature (spin-momentum locking), i.~e.~the two spin species of the edge modes propagate in opposite directions~\cite{Bernevig2006,Kane2005}.
Furthermore, the edge states are topologically protected ensuring robustness against perturbations with very long ($\gtrsim 1\mu m$) decoherence lengths~\cite{Chen2012,Molenkamp2007,Molenkamp2009,wiedenmann2016}.
These properties make TIs promising platforms for the manipulation of spin-entangled electrons in solid state systems. 

In this paper we demonstrate that combining s-wave superconductivity with the helical properties of 2D TIs \cite{jacquet2018}, the non-local manipulation of spin-entangled states by means of local gating can be done. The proposed setup [see Fig.~\ref{3D}(b)] is composed of two parallel 2D TIs properly connected to two superconducting electrodes, and comprises electrical gates for the manipulation.
We calculate analytically the current-phase relationship (CPR) of the Josephson current making use of the scattering matrix approach and we identify the various local and non-local scattering mechanisms. In particular, we show how the different external potentials selectively operate over the local and non-local components. We demonstrate that the application of gates affects the symmetry of the non-local entangled states (from singlet to triplet) which arise from crossed Andreev reflection between the two edges. This entanglement symmetry manipulation does not affect the purity of the entangled state but directly impact the Josephson coupling due to the intrinsic singlet nature of s-wave superconducting leads.   
We find that the Josephson critical current, remarkably, allows a direct quantification of the entanglement manipulation in the structure.
We fully interpret the described phenomenology in terms of the multiple Andreev processes which mediate the Josephson coupling in the structure. 

The paper is organized as follows. In Section~\ref{setup} we discuss the setup of the Josephson nanojunction done with topological insulators, we clarify why non-local entanglement may be realised and how the external potentials may be used to manipulate over the system. In Section~\ref{model} we briefly discuss how the computation in the scattering matrix formalism is done introducing the concept of electron losses at the TI-superconductor interfaces (all the details are reported in Appendices). In Section~\ref{field-action} we discuss euristically how external potentials affect the entanglement symmetry of the Cooper pairs in the junction, showing a simple way to interpret the complex behavior of the Josephson current. In Section~\ref{Josephson_current_section} we present our analytical and numerical results for the CPR and for the critical current, discussing with care the interpretation in terms of multiple Andreev reflection processes in some notable limits.
Finally we discuss the experimental feasibility of the proposal in the Conclusions.

\section{The Setup}
\label{setup}
In a Josephson system with ideal interfaces and rigid boundary conditions the phase difference $\phi=\phi_R-\phi_L$ between the two superconductors induces a stationary Josephson current. Microscopically it originates from Andreev reflection processes that 
describe the transfer of Cooper pairs (CPs) at the interfaces between the superconductors and the weak link.
In a single 2D TI sandwiched between two conventional s-wave superconductors [namely, the S-TI-S junction depicted in Fig.~\ref{3D}(a)], CPs can only be injected or absorbed locally on a specific edge. 
Indeed, while the helical nature of the TI edge modes allows for local Andreev reflections (LARs) at the boundaries with the superconductors \cite{Choi2014}, i.~e.~an electron (hole) propagating through a helical mode and impinging onto a superconductor is reflected as a hole (electron) with opposite spin in the other helical mode on the {\em same} edge [see right side of Fig.~\ref{3D}(a)],
it prohibits crossed Andreev reflections (CARs) \cite{crepin2015,fleckenstein2018}, i.~e. an electron (hole) propagating through a helical mode impinging onto a superconductor cannot be reflected as a hole (electron) with opposite spin in the other helical mode on the {\em other} edge [see left side of Fig.~\ref{3D}(a)].
In order to overcome this limitation one needs to consider a double TI junction \cite{Sato2010}. Specifically 
we focus on the architecture depicted in Fig.~\ref{3D}(b) where 
 a  Josephson junction is obtained by sandwiching two planes of 2D TIs in between two s-wave superconductors.
This system allows for CAR processes if the distance $W$ between the two TI-planes is comparable with the coherence length $\xi$,
e.~g. choosing Al as a superconducting material $\xi\approx 100$ nm.
Moreover, the properties of the edge modes can be tuned through external voltage gate and magnetic flux controls which mimics the presence of 
``local'' (i.e. which acts differently on the upper and lower edge modes) time reversal and time reversal breaking fields respectively [see Fig.~\ref{3D}(b)].

Specifically the first consists in gate electrodes placed in the vicinity of the edges modes (see Fig.~\ref{3D}(b)) so to electrostatically affect the dynamical phase of the carriers~\cite{xiao2016} by assigning to particles the same phase factor $\theta_{\rm V}=2eVL/\hbar v_F$ independently of their propagation direction and spin (with $v_F$ the Fermi velocity).
The second instead, consists in the application of a moderate, uniform magnetic field $B$ which, by Doppler shift effect~\cite{Tkachov2015,sothmann2017}, 
acts on the system by assigning a phase factor $\theta_{\rm \Phi}= 4\pi \Phi/\Phi_0$ to spin-up electrons and $-\theta_{\rm \Phi}$ to spin-down electrons (with $\Phi_0$ the magnetic flux quantum and $\Phi=B_y W L$ the magnetic flux in the junction).

The way we implement both the fields ensures their differential action among the upper and lower edges of the TIs such that they can be effectively described, in the spin space, in terms of local unitary operators:
\begin{equation}
\label{UV}
\mathcal{U}^{\rm \ell}_{\rm V}(\theta_{\rm V})= e^{i~\theta_{\rm V}\mathds{1}/2}
\end{equation}
\begin{equation}
\label{UPhi}
\mathcal{U}^{\rm \textit{u}}_{\rm \Phi}(\theta_{\rm \Phi})= e^{i~\theta_{\rm \Phi}\vb*{\sigma}\vdot\vu{n}/2}
\end{equation}
with $\vu{n}$ the natural spin-quantization axis of both the TIs~\cite{Qi2008,Maciejko2010} and with $\vb*{\sigma}$ the Pauli matrix vector.
In  Eqs.~($\ref{UV}$,~$\ref{UPhi}$) the labels $\textit{u}$ and $\ell$ indicate the action of the fields on the upper and lower edge respectively; other configurations, though, are fully equivalent as discussed in Appendix \ref{Appendix_D}.

%

\section{Model}
\label{model}
Following the scattering approach~\cite{Buttiker1986,Buttiker1992,Datta1997,Lambert1998}, 
the scattering matrix of the Andreev processes occurring on the left L (right R)  TI-S interface,  in the $u$-$\ell$ space, can be written as~\cite{beenakker1991,beenakker1992}
\begin{equation}
\begin{pmatrix}
\label{matrix}
\abs{\Lambda_{\rm L(R)}} & i \abs{X_{\rm L(R)}}\\ 
i \abs{X_{\rm L(R)}} & \abs{\Lambda_{\rm L(R)}}
\end{pmatrix}e^{i\phi_{\rm L(R)}},
\end{equation}
with  $\Lambda_{\rm L(R)}$ and $X_{\rm L(R)}$  representing respectively  the amplitude for the  LAR and  CAR events (these terms being related by the unitarity conditions $|\Lambda_{L(R)}|^2+|X_{L(R)}|^2=1$). 
In other words the Andreev reflection is given by the superposition of LAR and CAR processes.
Notice that we neglected the presence of the edge modes running in the backside edges (i.~e. along the $y$-direction) of the device of Fig.~\ref{3D}(b), which in turn represent an alternative coherent path between the two superconducting leads. Anyway one can nullify their contribution to the Josephson current by simply suppressing their transport coherence. This can be obtained assuming that the size of the TI in the $y$-direction is much larger than the coherence length $\ell_\phi$ or having intentionally broken superconducting coherence introducing a dephasing source along those edges~\footnote{This can be done by adding a floating metal pad over those edge modes which will induce electron decoherence.}. 
In such case they no longer contribute coherently to the transport even if they still represent available channels for ordinary reflected particles at interfaces. 
Such ordinary reflections, involving the backside modes, contribute by decreasing the supercurrent. This can be described by introducing an effective loss parameter $\eta\in[0,1]$ where $\eta=0$ represents lossless regime. When this loss mechanism is present, in order to describe the dephasing along the backside modes, we calculate the current by averaging with respect the dephasing angles -- see Appendix~\ref{The_Scheme} for details.





\section{Local fields selective action}
\label{field-action}
The $V$-field and $\Phi$-field defined before and respectively associated to the angles $\theta_{\rm V}$ and $\theta_{\rm \Phi}$, operate independently and selectively on the local and non-local components of the Josephson current~\footnote{The angular notation for the action of the local fields imply a $2\pi$-periodicity of their actions.}.
Before calculating explicitly the Josephson current in the model,  
a preliminary evidence of this fact is obtained via an heuristic argument applied to the simplified scenario where LARs are absent (i.~e. $\Lambda_{\rm L(R)}=0$). Under this circumstance the non-local emission of a CP from a superconducting electrode, say S$_{\rm L}$, results in the formation of a spin-entangled CP state, which arises from two superimposed CAR processes. In the first one, a spin-$\downarrow$ hole propagating in the lower edge gets reflected into an spin-$\uparrow$ electron in the upper edge, while in the second one, a spin-$\downarrow$ hole propagating in the upper edge gets reflected into an spin-$\uparrow$ electron in the lower edge [see Fig.~\ref{3D}(b)].
Such spin-entangled state could be represented as $\ket{C}=\left(\ket{e_u^{\uparrow} h_{\ell}^{\downarrow}}-\ket{h_u^{\downarrow}e_{\ell}^{\uparrow}}\right)/\sqrt{2}$,
where the minus sign recall the fact that the CP is in a spin-singlet state as required by the s-wave nature of the superconducting leads.
\begin{figure}[h!!]
\centering
    \includegraphics[width=.48\textwidth]{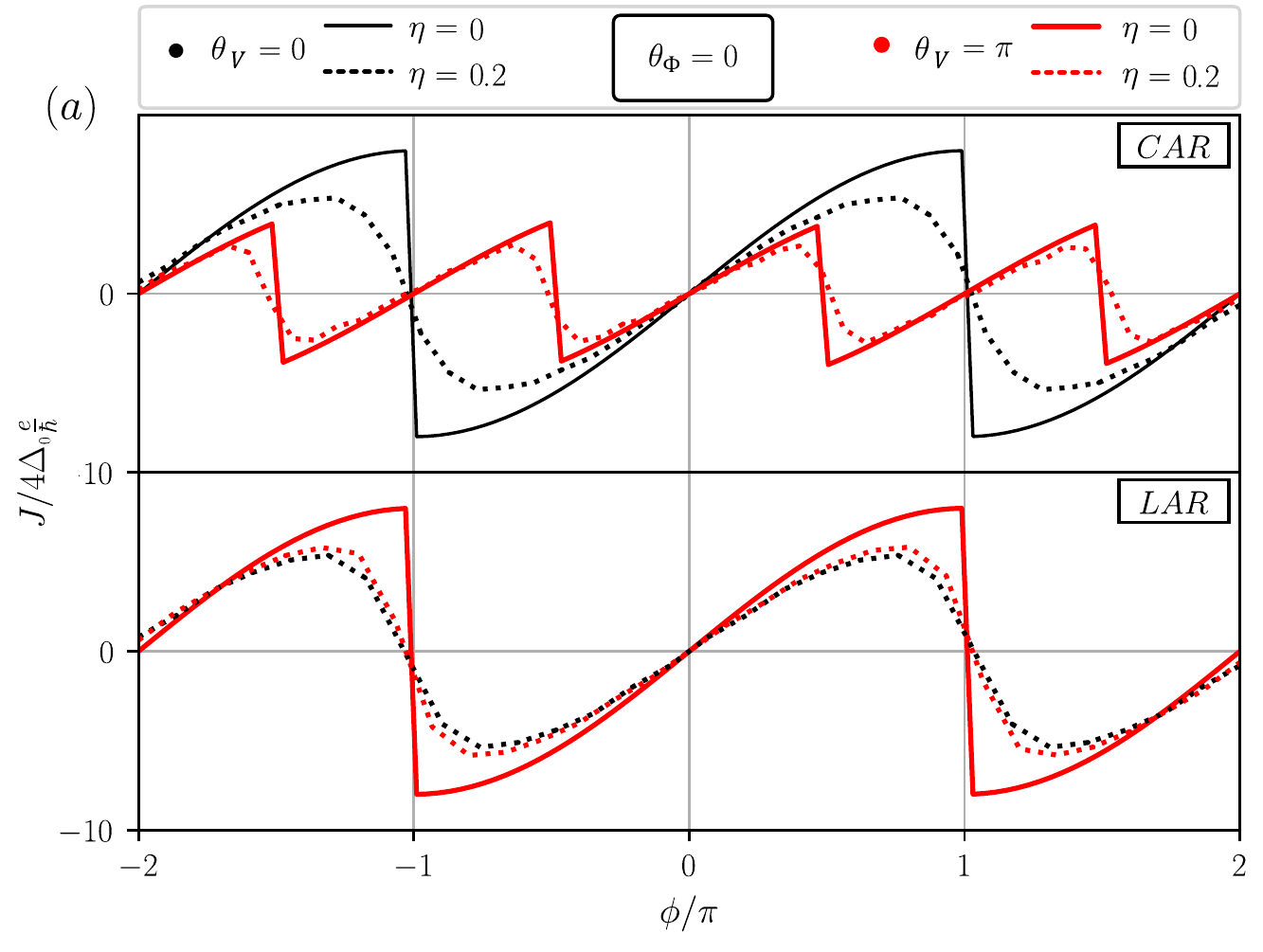}
    \includegraphics[width=.48\textwidth]{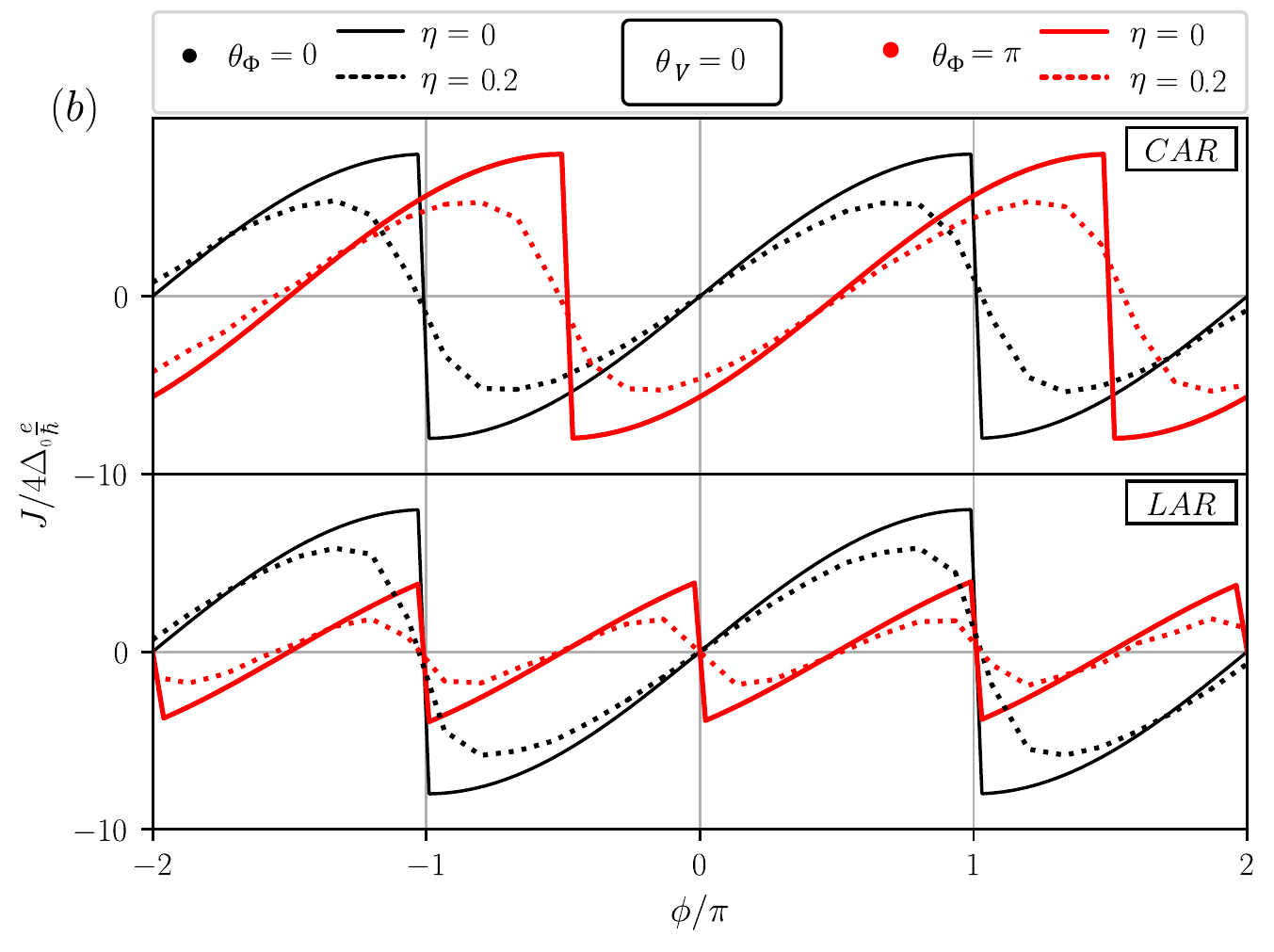}
    \caption{CPR. The Josephson current vs the phase difference $\phi$ expressed in units of $4\Delta_0\frac{e}{\hbar}$ at fixed temperature $k_BT/\Delta_0=10^{-3}$. The curves in panel $(a)$ have been obtained fixing the value of the magnetic field ($\theta_{\Phi}=0$), for different values of the $V$-field ($\theta_{\rm V}=0$ for the solid black line and $\theta_{\rm V}=\pi$ for the solid red line). Panel $(b)$ shows the opposite situation in which the $V$-field is fixed ($\theta_{\rm V}=0$) and $\theta_{\rm \Phi}$ is varied ($\theta_{\rm \Phi}=0$ for the solid black line and $\theta_{\rm \Phi}=\pi$ for the solid red line). The respective black/red dashed curves, depicting the case with $\eta=0.2$, have been obtained numerically. For both the realizations we have considered the two extremal cases, i.~e. only CAR $|X_{L(R)}|^2=1$ (upper panels) and only LAR $|X_{L(R)}|^2=0$ (lower panels).} 
    \label{Josephson_Current} 
\end{figure}
The action of $\mathcal{U}_{\rm V}(\theta_{\rm V})$ and $\mathcal{U}_{\rm \Phi}(\theta_{\rm \Phi})$
on $\ket{C}$ results in the state: $e^{i\frac{\theta_{\Phi}}{2}}\left(e^{-i\theta_{\rm V}/2}\ket{e_u^{\uparrow} h_{\ell}^{\downarrow}}-e^{+i\theta_{\rm V}/2}\ket{h_u^{\downarrow}e_{\ell}^{\uparrow}}\right)/\sqrt{2}$. This expression shows that while the $\Phi$-field introduces only a global phase, that can be reabsorbed with a gauge transformation, the $V$-field modifies the entanglement symmetry of the non-local CP state $\ket{C}$ by introducing a relative phase factor $\exp(i~\theta_{\rm V})$, without altering its entanglement content.
In particular, if $\theta_{\rm V}=\pi$ the non-local spin-singlet CP changes into a spin-triplet one, thus giving rise to a mismatch with respect to the intrinsic CPs singlet symmetry of the electrodes, thus hindering the Josephson coupling. In view of this fact, in the absence of LAR processes, one hence expects the Josephson current to depend upon the quantity 
${\cal C}=|\bra{C}\mathcal{U}_{V}(\theta_{V})\ket{C}|=|\cos(\theta_{V}/2)|$, which measures the degree of change of the symmetry of the entangled CP.
This emerges clearly from the study of the critical current, especially in the scenario where
 multiple Andreev reflections can be neglected (single-shot limit) -- see Eq.~(\ref{JCSINGLE}) below. 
The interplay between CAR and LAR and the possibility of 
multiple reflections, on the contrary, tend to reduce the visibility of the effect:  still, as we shall show in the following, also in this case the Josephson current keeps record 
of  the phenomenon in a way that ultimately allows  us to discriminating between CAR and LAR processes.

\begin{figure}[h!!]
\centering
    \includegraphics[width=.48\textwidth]{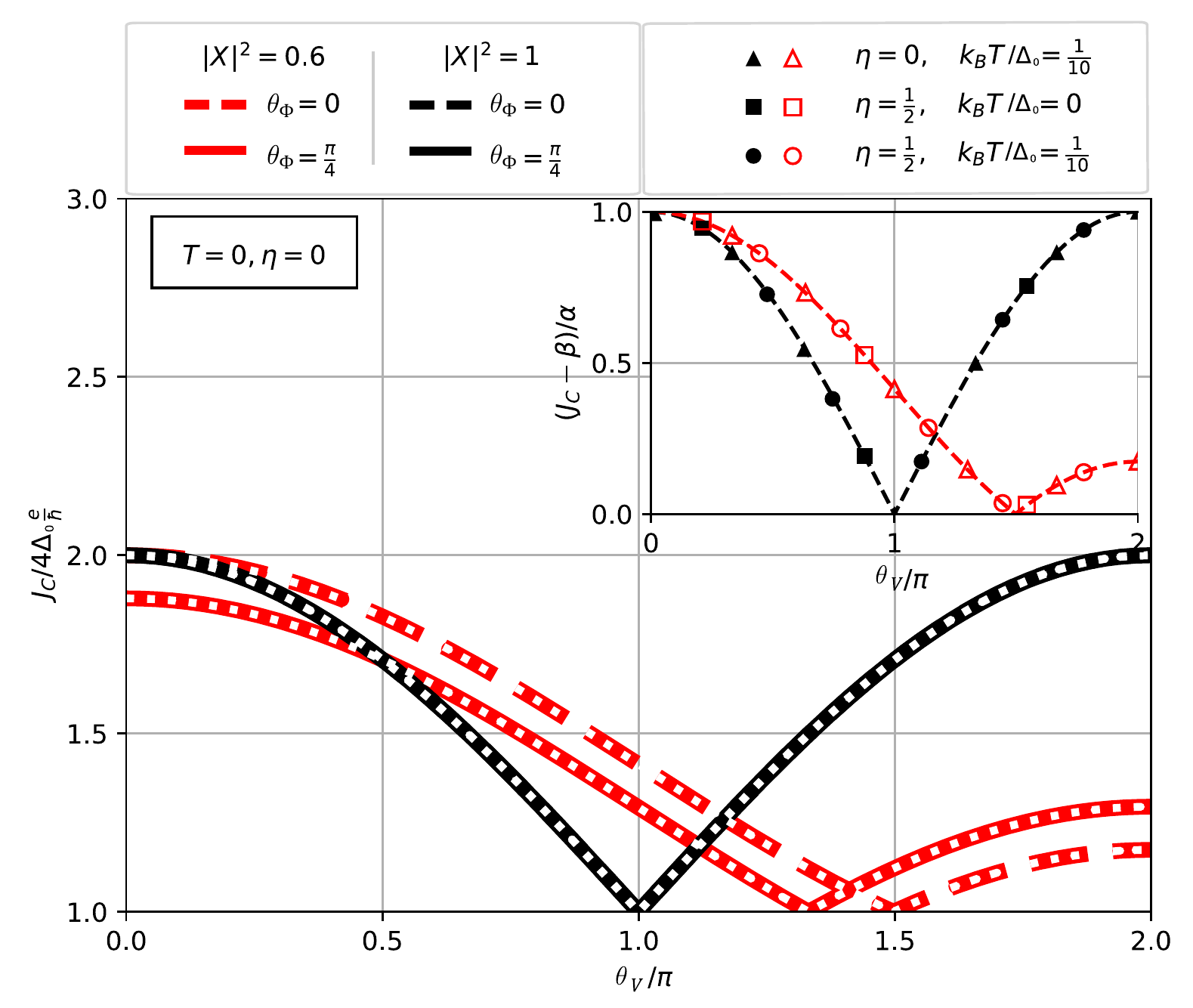}
    \caption{Critical current in units of $4\Delta_0\frac{e}{\hbar}$ as a function of $\theta_{V}$. 
In the main box, different set of black ($\abs{X}^2=1$) and red ($\abs{X}^2=0.6$) curves have been obtained numerically for fixed $T=0$ and $\eta=0$. 
For each $\abs{X}^2$, plots are shown in case of $\theta_{\rm \Phi}=0$ (dashed lines) and $\theta_{\rm \Phi}=\frac{\pi}{4}$ (solid lines) respectively. 
We superimposed (white dotted line) the analytical behaviour as predicted by Eq.~\eqref{EME_Critical_Current_main}.
The inset in the upper-right corner depicts the rescaled critical current (namely $(J_C-\beta)/\alpha$), obtained numerically for fixed $\theta_{\rm \Phi}=0$ in three different conditions of $\eta$ and $T$ (different point shape) for $\abs{X}^2=1$ (black color) and $\abs{X}^2=0.6$ (red color). All the data, independently of temperature and losses, perfectly match the curve $\abs{\Gamma}$ (dashed lines), confirming the universality character of the shape of the critical current upon different external parameters.} 
    \label{Critical_Fig} 
\end{figure}

\section{Josephson current}
\label{Josephson_current_section} To set the above observations on firm ground in the remaining of the paper we 
calculate the Josephson current flowing through the system using the scattering formalism  \cite{beenakker1991,beenakker1992,Beenakker2003} in the short junction limit (i.~e.~when $L\ll\xi$), with ideal interfaces and rigid boundary conditions, i.~e. with the order parameter $\Delta(x)=\Delta_0 e^{i\phi_L\theta(-x-L/2)+i\phi_R\theta(x-L/2)}$ and $\Delta(x)=0$ for $|x|\le\frac{L}{2}$, where $\theta(x)$ is the Heaviside function. 

The current can be calculated as $I=-\frac{2e}{\hbar}\sum_{p}\tanh\qty(\epsilon_p/2k_BT)\frac{d\epsilon_p}{d\phi}$,
where $\epsilon_p$ are Andreev bound state energies obtained solving the self-consistent secular problem~\cite{Beenakker2003}.
In case of no losses ($\eta=0$) we find the following analytical expression of the CPR at finite temperature~$T$  
\begin{align}
\label{J_current_analytical}
&J(\phi)= 4\frac{e\Delta_0}{\hbar}\sum_{\sigma=\pm}\Bigg\{\sin{\left(\frac{\theta_{\rm \Phi}}{4} + \frac{\phi}{2}+\sigma \tan^{-1}{\left(\sqrt{\frac{1-\Gamma_{}}{1+\Gamma_{}}}\right)}\right)}\times\nonumber\\&\tanh{\qty[\frac{\Delta_0}{2k_BT}\cos{\left(\frac{\theta_{\rm \Phi}}{4}+\frac{\phi}{2}+\sigma\tan^{-1}{\left(\sqrt{\frac{1-\Gamma_{}}{1+\Gamma_{}}}\right)}\right)}]}\Bigg\} ,
\end{align}
with 
\begin{equation}
\label{Gamma_main}
\Gamma=\cos{\left(\theta_{\rm V}/2\right)}\abs{X_L}\abs{X_R}+\cos{\left(\theta_{\rm \Phi}/2\right)}\abs{\Lambda_L}\abs{\Lambda_R}.
\end{equation}
Firstly we note that $\theta_{\rm \Phi}/4$ in Eq.~(\ref{Gamma_main}) acts as a global phase which shifts the CPR, and manifests itself as an anomalous current at $\phi=0$ when TRS is broken \cite{Dolcini2015,campagnano2015,minutillo2018}. For simplicity, in what follows we will limit ourselves to consider the fully symmetric case $\abs{X_{\rm L}}=\abs{X_{\rm R}}=\abs{X}$ and 
$\abs{\Lambda_{\rm L}}=\abs{\Lambda_{\rm R}}=\abs{\Lambda}=\sqrt{1-\abs{X}^2}$.

In Fig.~\ref{Josephson_Current}(a) and (b) we plot separately the contributions to the CPR arising from only CAR ($\abs{X}^2=1$ for top panels) and only LAR processes ($\abs{X}^2=0$ for bottom panels), for various choices of parameters.
Solid curves refer to the CPR of Eq.~(\ref{J_current_analytical}), while dashed curves are numerical results obtained in the presence of backside edges losses ($\eta=0.2$) \footnote{We tested that numerically for $\eta\to 0$ one obtain again the analytical result.}.
Firstly we note that the curves resembles the CPR of a weak-link in the presence of spin-orbit and magnetic fields, as we may indeed naively expect \cite{Yokoyama2013,Eto2014,Dolcini2015,Marra2016,Nava2016,Mellars2016,Nesterov2016}. 
In Fig.~\ref{Josephson_Current}(a) we fix $\theta_{\rm \Phi}=0$ and consider two values of the $V$-field, namely $\theta_{\rm V}=0$ (black curves) and $\theta_{\rm V}=\pi$ (red curves), while in Fig.~\ref{Josephson_Current}(b) we fix $\theta_{\rm SO}=0$ and consider two values of  $\Phi$-field, namely $\theta_{\rm \Phi}=0$ (black curves) and $\theta_{\rm \Phi}=\pi$ (red curves). 

Figs.~\ref{Josephson_Current}(a) and (b) allows us to appreciate the selective action of the $\Phi$- and $V$- fields on the CAR and LAR contributions to the supercurrent by their effect on the shape of the CPR. 
In particular, in the case of CAR processes, the shape of the CPR depends on the value of $\theta_{\rm V}$ [Fig.~\ref{Josephson_Current}(a), top panel], independently of the value of $\theta_{\rm \Phi}$, which only induces a global phase-shift [Fig.~\ref{Josephson_Current}(b), top panel]. Indeed in Fig.~\ref{Josephson_Current}(a), where we fix for simplicity $\theta_{\rm \Phi}=0$, black ($\theta_{V}=0$)  and red ($\theta_{\rm V}=\pi$) curves 
have different shapes in the top panel ($\abs{X}^2=1$ only CAR) differently to the bottom panel ($\abs{X}^2=0$ only LAR) where the black and red curves are superposed having exactly the same shape.
Conversely, as shown in Fig.~\ref{Josephson_Current}(b), the CPR shape is affected by the value of $\theta_{\rm \Phi}$ (black lines $\theta_{\rm \Phi}=0$ and red ones $\theta_{\rm \Phi}=\pi$) when only LAR processes are present (bottom panel) but not affected, forgetting an unessential global phase shifting, in case of only CAR processes. For other values of $\theta_{\rm V}$ the CPR shape is changed, in comparison to the figure $\theta_{\rm V}=0$, but the shape changes with $\theta_{\rm \Phi}$ only when LAR contribution are indeed present. 

We can conclude that, although in general the Josephson current contains both CAR and LAR contributions, any variation of the shape of the CPR due to the action of the $V$-field is a direct indication of the presence of CAR processes, (i.~e.~of non-local injection of spin-singlet CPs).

The presence of losses ($\eta\neq 0$) simply leads to a smoothing of the CPR shape, similarly to the effect of a finite temperature~(see section~$\ref{finite_Temperature_and_eta}$), but not affect the previous discussion.

\subsection{Single-shot limit}\label{single_shot_regime}
At this point is interesting to investigate the behaviour of the Josephson current in the limit of high losses where $\eta\approx 1$, i.~e. looking at the lowest order in $(1-\eta)$. In that regime few CPs tunnel into the junction (the Josephson current is dramatically reduced). Different orders in the power $(1-\eta)$ corresponds to bounces of Cooper pairs at the interfaces with the superconductors, i.~e. multiple Andreev processes. In particular one expects that the lowest order of the Josephson current corresponds to a term proportional to $(1-\eta)^4$ which describes a process where a Cooper pair is emitted on one side and absorbed on the other side and viceversa~\footnote{The lowest order $(1-\eta)^4$ accounts for the single shot CP process, where the CP is splitted at one barrier, taking an $(1-\eta)^2$ factor, and another factor when it recombines on the other barrier}. This corresponds to the propagation of the single
Cooper pair along the junction (single shot limit). Higher orders $(1-\eta)^\alpha$ with $\alpha>4$ corresponds to multiple Andreev processes where the emitted Cooper pair is at least reflected back one time. 

We present below the analytical results of the Josephson current in the single shot limit following the scheme presented in the previous section.
More specifically, we considered two scenarios (the same as those represented in Fig.~\ref{Josephson_Current}~(a)-(b)):
in the first case we just account for the local application of the $V$-field along one of the edge states of the TI (hence setting $\theta_{\rm \Phi}=0$); while in the second case the $\Phi$-field is applied by fixing $\theta_{\rm V}=0$. 
Again, per each scenario, we consider the extremal situations in which only CAR ($\abs{X_L}=\abs{X_R}=\abs{X}=1$) or only LAR ($\abs{X}=0$) processes are involved at both the interfaces, in order to have only non-local or local CP-splitted states inside the junction.

\begin{itemize}
\item Case-(a): application of $V$-field ($\theta_{\rm \Phi}=0$)
\end{itemize}
\begin{subequations}
\label{SO_term}
\mybox{CAR}~:
\begin{align}
\label{SO_a}
\bar{J}(\phi)=&\frac{e}{\hbar}\Delta_0\tanh{\left(\frac{\Delta_0}{2T}\right)}\cos{\left(\frac{\theta_{\rm V}}{2}\right)}\sin{\left(\phi\right)}(1-\eta)^4\nonumber\\&+\mathcal{O}((1-\eta)^{6}),
\end{align}
\mybox{LAR}~:
\begin{align}
\label{SO_b}
\bar{J}(\phi)=&\frac{e}{\hbar}\Delta_0\tanh{\left(\frac{\Delta_0}{2T}\right)}\sin{\left(\phi\right)}(1-\eta)^4\nonumber\\&+\mathcal{O}((1-\eta)^{6}).
\end{align}
\end{subequations}

\begin{itemize}
\item Case-(b): application of $\Phi$-field ($\theta_{\rm V}=0$) 
\end{itemize}
\begin{subequations}
\label{B_term}
\mybox{CAR}~:
\begin{align}
\label{B_a}
\bar{J}(\phi)=&\frac{e}{\hbar}\Delta_0\tanh{\left(\frac{\Delta_0}{2T}\right)}\sin{\left(\phi+\frac{\theta_{\rm \Phi}}{2}\right)}(1-\eta)^4\nonumber\\&+\mathcal{O}((1-\eta)^{6}),
\end{align}
\mybox{LAR}~:
\begin{align}
\label{B_b}
\bar{J}(\phi)=&\frac{e}{\hbar}\Delta_0\tanh{\left(\frac{\Delta_0}{2T}\right)}\left[\sin{\left(\phi\right)}+\sin{\left(\phi-\frac{\theta_{\rm \Phi}}{2}\right)}\right](1-\eta)^4\nonumber\\&+\mathcal{O}((1-\eta)^{6}).
\end{align}
\end{subequations}

Here $\bar{J}(\phi)$ represents the CPR averaged with respect the dephasing angles acquired along the backside edges of the model -- see Appendix~\ref{The_Scheme}.
It is important to note that Eqs.~(\ref{SO_term},~\ref{B_term}) are fully in agree with the behaviours of the  Josephson current as described by the results reported in Fig.~\ref{Josephson_Current}. In particular, in Case-(a), the $V$-field just affects the CAR component of the CPR Eq.~(\ref{SO_a}) while plays no role when only LAR processes are involved Eq.~(\ref{SO_b}).
In Case-(b), the action of the $\Phi$-field just operates as a global shifting on the CAR component of the CPR Eq.~(\ref{B_a}), while it affects the shape of the supercurrent in case only LAR processes occur at both the interfaces Eq.~(\ref{B_b}). It is worth noting that the modification of the shape of the CPR in the cases related to Eqs.~(\ref{SO_a},~\ref{B_b}) respectively, have two different origins. More specifically, in the case of Eq.~(\ref{B_b}), as a consequence of the presence on only LAR processes the two CPs are separately injected in the two different TI-planes. The resulting Josephson current takes the form of a sum of two independent currents: one concerning the edge of the TI interested by the application of the $\Phi$-field (which is shifted by an amount of $\theta_{\rm \Phi}/2$), and the other, that instead refers to the free edge, which takes the usual form of $\sin{(\phi)}$. On the contrary, for only CAR the modification of the profile of the CPR of Eq.~(\ref{SO_a}), which is ruled by the factor $\cos{(\theta_{\rm \Phi}/2)}$, exactly reflects the action the $V$-field which operates on the non-local states affecting their entanglement symmetry as discussed in~section~\ref{field-action}. For the only CAR case this is even clearer by looking at the critical current of the system as we will see in the next section.

\subsection{Critical current}\label{critical_current_main} Let us now consider the behaviour of the critical current, defined as $J_{\rm c}= \max_{\phi} \left\{\left|J(\phi)\right|\right\}$, which is plotted in the main panel of Fig.~\ref{Critical_Fig} as a function of $\theta_{\rm V}$ for different sets of parameters.
Remarkably, we find that the critical current can be written in the following compact form 
\begin{equation}
\label{EME_Critical_Current_main}
J_{\rm c}=\alpha(\eta,T)\left|\Gamma\right|+\beta(\eta,T),
\end{equation}
where $\Gamma$, defined in Eq.~\eqref{Gamma_main}, depends only on the Andreev reflection amplitudes $X_L$ and $X_R$, and on the fields strengths $\theta_{\rm \Phi}$ and $\theta_{\rm V}$, while the prefactor $\alpha$ and the off-set $\beta$ depend only on the temperature $T$ and on the losses $\eta$.

The main panel of Fig.~\ref{Critical_Fig} shows how the formula of Eq.~\eqref{EME_Critical_Current_main} (white dotted lines) exactly fits the numerical results of the critical current (black and red lines) for an arbitrary choice of the CAR/LAR amplitudes and of the manipulation parameters. We first consider the ideal case of no losses $\eta=0$ and $T=0$ for which $\alpha=\beta=1$. The different lines corresponds to different cases: only CAR $|X|^2=1$ (black lines) and the intermediate case with CARs and LARs both present $|X|^2=0.6$ (red lines). We show with solid lines the cases $\theta_{\rm \Phi}=0$ and with dashed lines $\theta_{\rm \Phi}=\pi/4$.

We see, for only CAR, that the minimum of $J_c$ occurs at $\theta_{\rm V}=\pi$ and $\theta_{\rm \Phi}$ does not affect $J_c$ (solid and dashed curves coincides), in full agreement with the discussion done before on the CPR. Red lines shows that for the case where both CAR and LAR contributions are present the $J_c$ is still described by the general formula for any value of $\theta_{\rm \Phi}$.

Furthermore we can show the general validity of this formula for finite values of $\eta$ and $T$. In the inset of Fig.~\ref{Critical_Fig} we plot the quantity $(J_C-\beta)/\alpha$ for different values of the temperatures and losses (see label). All the points perfectly match the corresponding $\abs{\Gamma}$ curve (thin dashed lines) as predicted by Eq.~\eqref{EME_Critical_Current_main}.

Hereafter we claim that the dependence on $\theta_{\rm V}$ of $J_c$ such as the one shown by  Eq.~\eqref{EME_Critical_Current_main} reflects the entanglement symmetry manipulation due to the action of the $V$-field.
We first notice that the critical current, resulting from Andreev bound states within the junction, can be seen as consisting of the sum of contributions arising from multiple Andreev reflection processes. In the only-CAR regime one can identify, for any values of $\eta$, two classes of processes: the ones corresponding to Cooper pairs which traverse the junction back and forth an even number of times and the processes which traverse the junction an odd number of times.
For the even class, the singlet symmetry is not modified by the effect of the $V$-field, since the backward time-reversed propagation cancels the $V$-field induced phase taken during the forward propagation.
The spin entanglement symmetry is instead changed only for the odd class processes.
This suggests that, at zero temperature and without losses ($\eta=0$), the odd class processes contribute to the critical current with the term, introduced before, ${\cal C}=
|\bra{C}\mathcal{U}_{V}(\theta_{\rm V})\ket{C}|=|\cos(\theta_{\rm V}/2)|$ 
in units of $J_0=4\Delta_0\frac{e}{\hbar}$. At the same time the even class is independent of $\theta_{\rm V}$ and  contributes to the current with the constant value $J_0$ (this give rise to the off-set $\beta$ in Eq.~\eqref{EME_Critical_Current_main}).
As a result, the critical current can be written as $J_{\rm c}=J_0(1+\cal C)$.
In particular, at $\theta_{\rm V}=\pi$ the entanglement symmetry of the non-local electronic state is changed into triplet in half of the processes (the odd ones) and is left singlet in the other half (the even ones). As a result, the non-local electronic state is an equal weighted mixture of singlet and triplet states.
This interpretation is actually corroborated by the fact that when only the lowest order processes contribute, i.~e. in the single-shot limit occurring when $\eta\simeq 1$, the critical current in the leading term of ($1-\eta$) takes the form
\begin{eqnarray} J_c=\frac{e\Delta_0}{\hbar}\abs{\cos\left(\theta_{\rm V}/2\right)}(1-\eta)^4+\mathcal{O}((1-\eta)^6) \;, \label{JCSINGLE}\end{eqnarray} 
that is equal to zero in when $\theta_{\rm V}=\pi$.
Such a result shows that in the single shot regime the action of the local $V$-field returns exactly the expected entanglement manipulation signature ${\cal C}$. 
Furthermore, Eq.~\eqref{EME_Critical_Current_main} clarifies that the critical current allows one to access experimentally the product $|X_R| |X_L|$ which determines the relative weight between the LAR and CAR processes.
Ultimately this  can be seen as a consequence of the selective action of the fields  on the local and non-local components of the current (second and first term in $\Gamma$).

 \subsection{Effect of the temperature $T$ and $\eta$}\label{finite_Temperature_and_eta}
 
 In this section we investigate in more detail the effect of the temperature $T$ and losses $\eta$ on the critical current.
\begin{figure}[h!!]
\centering
    \includegraphics[width=.5\textwidth]{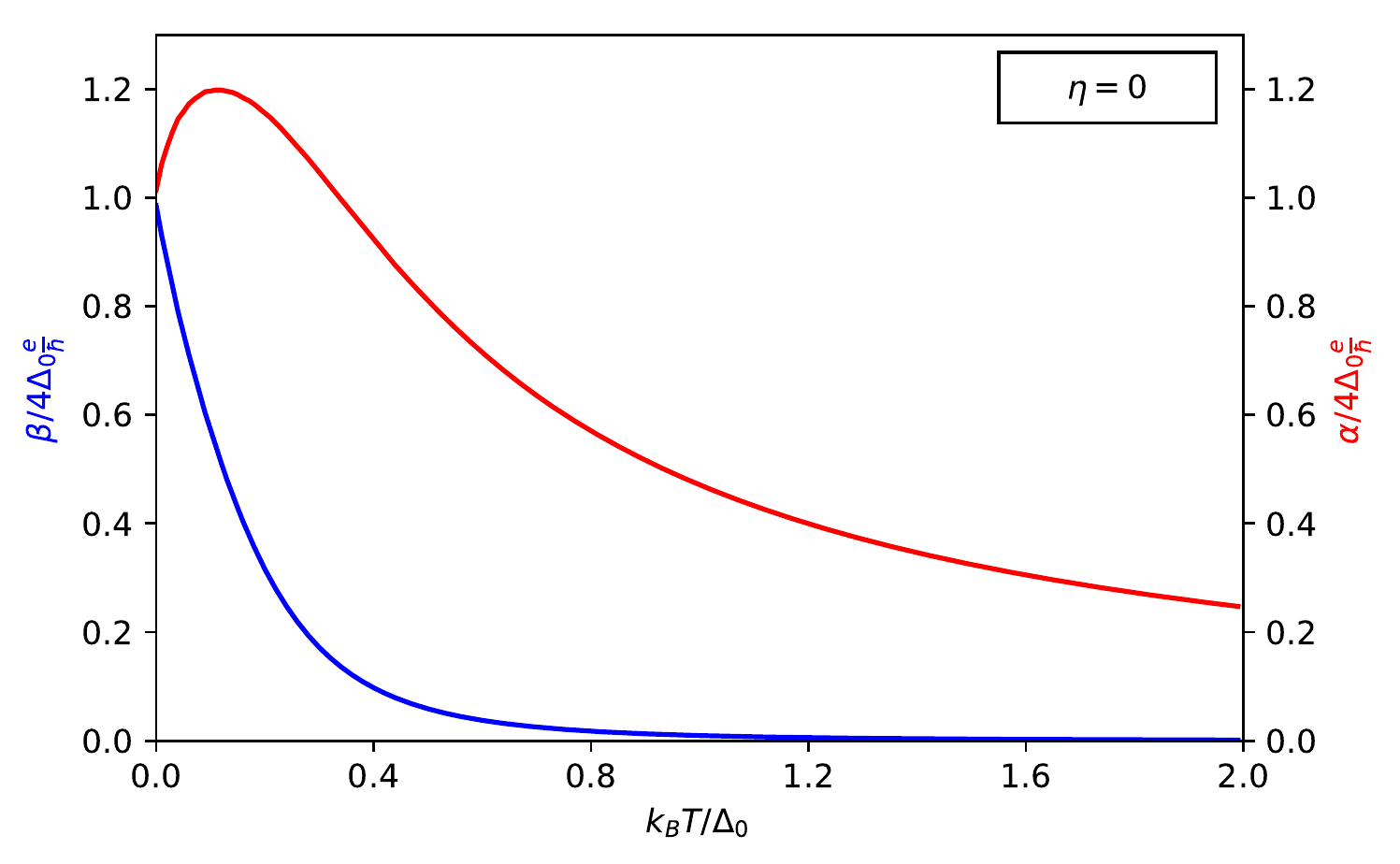}
   \caption{ Numerical plots of the amplitude $\alpha$ (red line) and the off-set $\beta$ (blue line) of the critical current of Eq.~($\ref{EME_Critical_Current_main}$) in case of no losses ($\eta=0$),  expressed in units of $4\Delta_0\frac{e}{\hbar}$ as functions of the thermal energy $k_BT$ (in units of $\Delta_0$). Note in the limits of $k_BT\ll \Delta_0$ and $k_BT\gg \Delta_0$ we recover the analytical results of Eq.~(\ref{Tchar}).}
    \label{alpha_beta} 
\end{figure}
As pointed out in section~\ref{critical_current_main}, the general expression of the critical current of the system takes the compact form of Eq.~\eqref{EME_Critical_Current_main}
where the amplitude $\alpha$ and the off-set $\beta$ just depend on the temperature $T$ and on the losses $\eta$. In Sec.~\ref{single_shot_regime} we calculate the Josephson current at the lowest order in $(1-\eta)$ for any temperature, see Eqs.~(\ref{SO_term},~\ref{B_term}).
$J_{\rm c}$ can be also calculated analytically in the regime of no losses ($\eta=0$) in the limit cases of low ($k_BT\ll\Delta_0$) and high ($k_BT\gg\Delta_0$) temperature. Indeed, we find
\begin{subequations}
\label{Tchar}
\begin{empheq}[left={\empheqlbrace\,}]{align}
\label{T0}
J_{\rm c}\simeq J_0(1+\abs{\Gamma(\theta_{V},\theta_{\Phi})}) & ~\text{for} ~ k_BT\ll\Delta_0\\
\label{Tinf}
J_{\rm c}\simeq J_0\frac{\Delta_0}{2k_BT}\abs{\Gamma(\theta_{V},\theta_{\Phi})}&~\text{for}~ k_BT\gg\Delta_0
\end{empheq}
\end{subequations}
which clearly shows the fundamental dependence of the critical current on the $\Gamma$ function of Eq.~\eqref{Gamma_main}.  
In particular, in the limit of low temperature, Eq.~(\ref{T0}), we have $\alpha=\beta=1$, while for high temperatures, Eq.~(\ref{Tinf}), we have $\alpha/J_0=\frac{\Delta_0}{2k_BT}$ and $\beta=0$. The numerical plot of $\alpha$ (red line) and $\beta$ (blue line) as functions of $k_BT$, in the case $\eta=0$, is shown in  Fig.~$\ref{alpha_beta}$. 

We note that by increasing the temperature, the critical current is depressed being both $\alpha$ and $\beta$ decreasing ($\beta$ gets suppressed much faster than $\alpha$). The behavior of the off-set $\beta$ as a function of temperature is consistent with the interpretation, given in section~\ref{critical_current_main}, that it corresponds to the contribution to the critical current of processes where CPs bounce back and forth along the junction, i.~e. multiple Andreev reflections. Indeed increasing the temperature we expect that multiple Andreev processes are strongly suppressed in comparison to the single transmission of a Cooper pair which will dominate the Josephson current contribution in the high temperature regime. This is why, in the high-temperature limit, the critical current is directly proportional to the $\Gamma$ function which effectively describes the manipulation induced by the local fields over a single CP transfer.

\begin{figure}[h!!]
\centering
    \includegraphics[width=.5\textwidth]{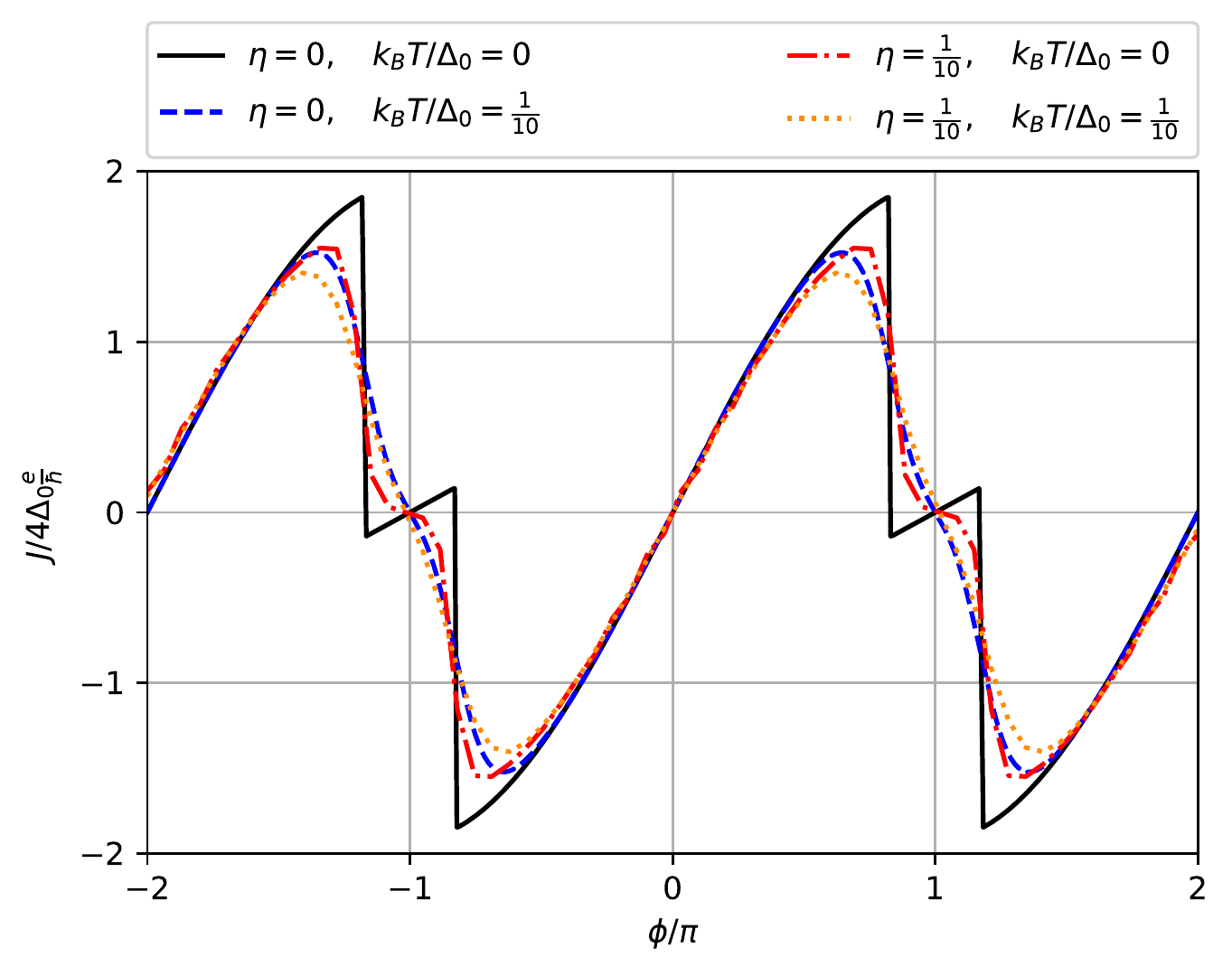}
    \caption{Josephson CPR expressed in units of $4\Delta_0\frac{e}{\hbar}$. For the sake of simplicity, symmetric conditions have been considered at left/right boundaries for the splitting amplitudes: $\abs{X_L}=\abs{X_R}=\abs{X}$. All the curves have been obtained for the following choice of parameters: $\abs{X}^2=1/2$, $\theta_{\rm \Phi}=0$, $\theta_{V}=\pi/2$. Solid lines correspond to the ideal case of no losses, i.~e. $\eta=0$, with $k_BT=0$ (dark-blue solid line) and $k_BT/\Delta_0=1/10$ (light-blue solid line). Dashed lines depict the case of finite losses, namely $\eta=1/10$, for the same values of temperature as before, i.~e. $k_BT=0$ (red dashed line) and $k_BT=1/10$ (magenta dashed line). The result of finite losses $\eta$ is to smooth out the shape of the Josephson current similarly to the effect of a finite temperature. } 
    \label{Critical_shape_T_eta}
\end{figure}

We discussed at the beginning of section~\ref{Josephson_current_section} that the presence of losses $\eta$ only smoothens the shape of the CPR.
In Fig.~$\ref{Critical_shape_T_eta}$ we compare the CPR in the case of finite $\eta$ (dashed lines) with the case of no-losses (solid lines) for two different  temperatures $T=0$ and $k_BT/\Delta_0=0.1$.
One can easily see how the smoothening induced by the losses described by $\eta$ are similar to the smoothening induced by the temperature effects.


\section{Conclusions} 
In this paper we have proposed a system which makes use of helical edge states of a 2D topological insulator (TI), in a specific configuration, to spatially separate the two electrons composing a Cooper pair of an s-wave superconductor. Such spatial separation can described, in the scattering approach, as a crossed Andreev reflection (CAR) process. The application of an external gate potential, which do not break time-reversal symmetry, enables the manipulation of the entanglement symmetry of the CAR state preserving its purity.
We have also shown that a time-reversal breaking field can be used to tune the strength of the local Andreev reflection (LAR) processes without affecting the previously discussed entanglement manipulation.     
In particular we have shown, both analytically and numerically, that a measurable signature of the manipulation is provided by the Josephson current. We have derived the analytical formula for the current-phase-relationship in the structure as a function of the external fields, for any temperature in the absence of losses $\eta=0$.   
In this configuration, the critical current can be directly connected to the relative weights between LAR and CAR processes, thus representing a simple way to identify the existence of non-local processes. Finally we have demonstrated, by carefully discussing the multiple Andreev reflections process occurring in the structure, the origin and the universality (independently of  temperature or losses) of the obtained results. In essence, the Josephson current phenomenology is naturally associated 
to the entanglement symmetry evolution in the junction. We think that the proposed structure, can be realized with present technology of the hybrid topological nanojunctions, thus opening a new route toward entanglement manipulation in electronic solid state systems. 

We conclude by estimating the strength of the potential voltage necessary to manipulate the entanglement symmetry from singlet to triplet.
In this case, in order to have $\theta_{\rm V}\approx \pi$ --- with a junction length of $L\approx 600$ nm~\cite{wiedenmann2016} and a TI Fermi velocity of $v_F=10^5$ m/s --- the gate voltage can be estimated as $V=1.7$ mV. A similar estimation shows that, using the Doppler shift,
the magnetic field strength necessary to perform a manipulation of the angle $\theta_{\rm \Phi}\approx \pi$ --- able to suppress the LAR component --- is $8$ mT, which is not too prohibitive in order to not break the topological protection of the TI.

\section{Acknowledgemnt}

We are grateful to Prof. M. Governale and Prof. L. Molenkamp for discussions. A.B. acknowledges the European
Research Council under the European Union’s Seventh Framework Program (FP7/2007-2013)/ERC Grant
agreement No. 615187-COMANCHE and the Royal Society though the International Exchanges between the UK and Italy (grant IESR3 170054). F. T. and A. B. the CNR-CONICET cooperation programme “Energy conversion in quantum nanoscale hybrid devices”.

\appendix

\section{Effective Hamiltonian}
The helical edge states at boundaries of each TI of the system are described by a one-dimensional Dirac Hamiltonian
\begin{align}
\label{Effective_edge_Hamiltonian}
H_{k}=&\hbar v_F \sum_{\zeta=\pm} \int dx\times\nonumber\\ &\left[\psi_{\zeta k\downarrow}^{\dagger}(\zeta i\partial_x-\mu)\psi_{\zeta k\downarrow}-\psi_{\zeta k\uparrow}^{\dagger}(\zeta i\partial_x+\mu)\psi_{\zeta k\uparrow} \right]
\end{align}
where $k=u,\ell$ labels the upper and lower TI plane, $\psi_{\uparrow}$ ($\psi_{\downarrow}$) is the field operator of $\uparrow$($\downarrow$) electrons, $\mu$ is the chemical potential and $v_F$ is the propagation Fermi velocity. The index $\zeta$ is associated to the front-side ($\zeta=+$) or backside ($\zeta=+$) edges, see Fig.~1(b) of the main text.
For the sake of simplicity we considered the same spin-quantization axis for both the TI planes edges along the $\vu{n}$ direction \cite{Qi2008,Maciejko2010}.
In the case this condition is not realized  one need to generalize our approach to the case of not collinear spin quantization axis. In this case some of the simple analytical results are not anymore valid, but numerically all the calculations can be repeated. Nonetheless the main results of the paper, such as the selective action of the $V$-field over the entanglement symmetry, are still valid since they are based purely on general symmetry arguments.
Furthermore not collinear natural spin quantization axis would potentially results also in a reduction of the CAR injection in favour of the LAR processes. 

\section{Gate potential}\label{gating_potential}

Let us consider the Schrödinger equation for the topological effective edge Hamiltonian 
\begin{widetext}
\begin{equation}
\label{Schrodinger_Eq}
 \mqty(-i\hbar v_F\partial_x+e V(x)-\epsilon_F & 0 \\ 0  &  i\hbar v_F\partial_x+e V(x)-\epsilon_F)\mqty(u(x)  \\ d(x) )=E\mqty(u(x)  \\ d(x) )
\end{equation}
\end{widetext}
in which we considered the application of a constant gate potential 
\begin{align*}V(x) =  \begin{cases}    \,\,0&  \quad  \text{for } \abs{x}>\frac{L}{2}\\    \,\,V&  \quad  \text{for } \abs{x}\le\frac{L}{2} \end{cases}
\end{align*}
In in Eq.~($\ref{Schrodinger_Eq}$) we expressed the wave function $\Psi=\mqty(u(x)  \\ d(x) )$ in spinorial notation for the spin-up and spin-down components (along the $\vu{n}$ natural spin-quantization axis), while $\epsilon_F$ and $v_F$ represent the Fermi energy and the Fermi velocity respectively. The general form of the solution is the following
\begin{align*}\Psi(x) =  
\begin{cases}    
\,\,A\mqty(1  \\0)e^{ikx}+B\mqty(0  \\1)e^{-ikx}&  ~  \text{for } x<-\frac{L}{2} ~ (I)\\  
\\  
\,\,C\mqty(1  \\0)e^{ik^{\prime}x}+D\mqty(0  \\1)e^{-ik^{\prime}x}&  ~  \text{for } \abs{x}\le\frac{L}{2}~ (II)\\
\\
\,\,F\mqty(1  \\0)e^{ikx}+G\mqty(0  \\1)e^{-ikx}&  ~  \text{for } x>\frac{L}{2}~ (III)
\end{cases}
\end{align*}
where, in the limit of low energies ($E\ll\epsilon_F$), $k\approx k_F$ and $k^{\prime}\approx k_F-\frac{e V}{\hbar v_F}$.
In order to obtain the relation between the coefficients A,B, $\dots$ ,G, one uses the continuity requirements for the wave function $\Psi$ and the current 
\begin{align}
\lim_{\delta\rightarrow0}\Psi(x)\mid_{-L/2-\delta}^{-L/2+\delta}=0,\quad\quad & \lim_{\delta\rightarrow0}\Psi(x)\mid_{L/2-\delta}^{L/2+\delta}=0\\\nonumber
\lim_{\delta\rightarrow0}\hat{J}\Psi(x)\mid_{-L/2-\delta}^{-L/2+\delta}=0, \quad\quad& \lim_{\delta\rightarrow0}\hat{J}\Psi(x)\mid_{L/2-\delta}^{L/2+\delta}=0
\end{align}
where $\hat{J}\equiv\frac{\partial\hat{H}}{\partial\hat{p}}=v_F \sigma_z$ is the current operator for the hamiltonian defined in Eq.~($\ref{Effective_edge_Hamiltonian}$) with $\sigma_z$ the $z$-Pauli matrix.
By following standard procedures \citep{Schwabl} one can calculate the transmission $t$ and reflection $r$ amplitudes through the region \textbf{II}, in the following cases:
\begin{itemize}
\item A particle incident from the left (region $\textbf{I}$), i.~e. $A=1$, $G=0$:
\begin{equation}
\label{t_I-III}
t_{I \rightarrow III}=\frac{F}{A}=e^{-iL\frac{eV}{\hbar v_F}}; \quad \quad r_{I \rightarrow I}=\frac{B}{A}=0
\end{equation}
\item A particle incident from the right (region $\textbf{III}$), i.~e. $A=0$, $G=1$:
\begin{equation}
\label{t_III-I}
t_{III \rightarrow I}=\frac{B}{G}=e^{-iL\frac{e V}{\hbar v_F}}; \quad \quad r_{III \rightarrow III}=\frac{A}{G}=0
\end{equation}
\end{itemize}
The only contribution of $V$ is to generate a dynamical phase in the electron propagation.
From Eqs.~($\ref{t_I-III},\ref{t_III-I}$) is clear that the constant potential barrier acts by assigning to electrons the same phase factor independently from their propagation direction and spin, 
clarifying why the unitary operator of Eq.~(\ref{UV}) takes the form $\mathcal{U}_{\rm V}(\theta_{\rm V})= e^{i~\theta_{\rm V}\mathds{1}/2}$, with $\mathds{1}$ the identity operator in the spin space. In particular, in this case $\theta_{\rm V}= 2e V L/(\hbar v_F)$. 
This phase indeed coincides with the dynamical phase acquired by an electron propagating along the edge under the electrical potential $V$ for a time of flight $t=L/v_F$ \cite{xiao2016}. 

\section{The scheme}\label{The_Scheme}

\begin{figure*}[ht]
\centering
    \includegraphics[width=.8\textwidth]{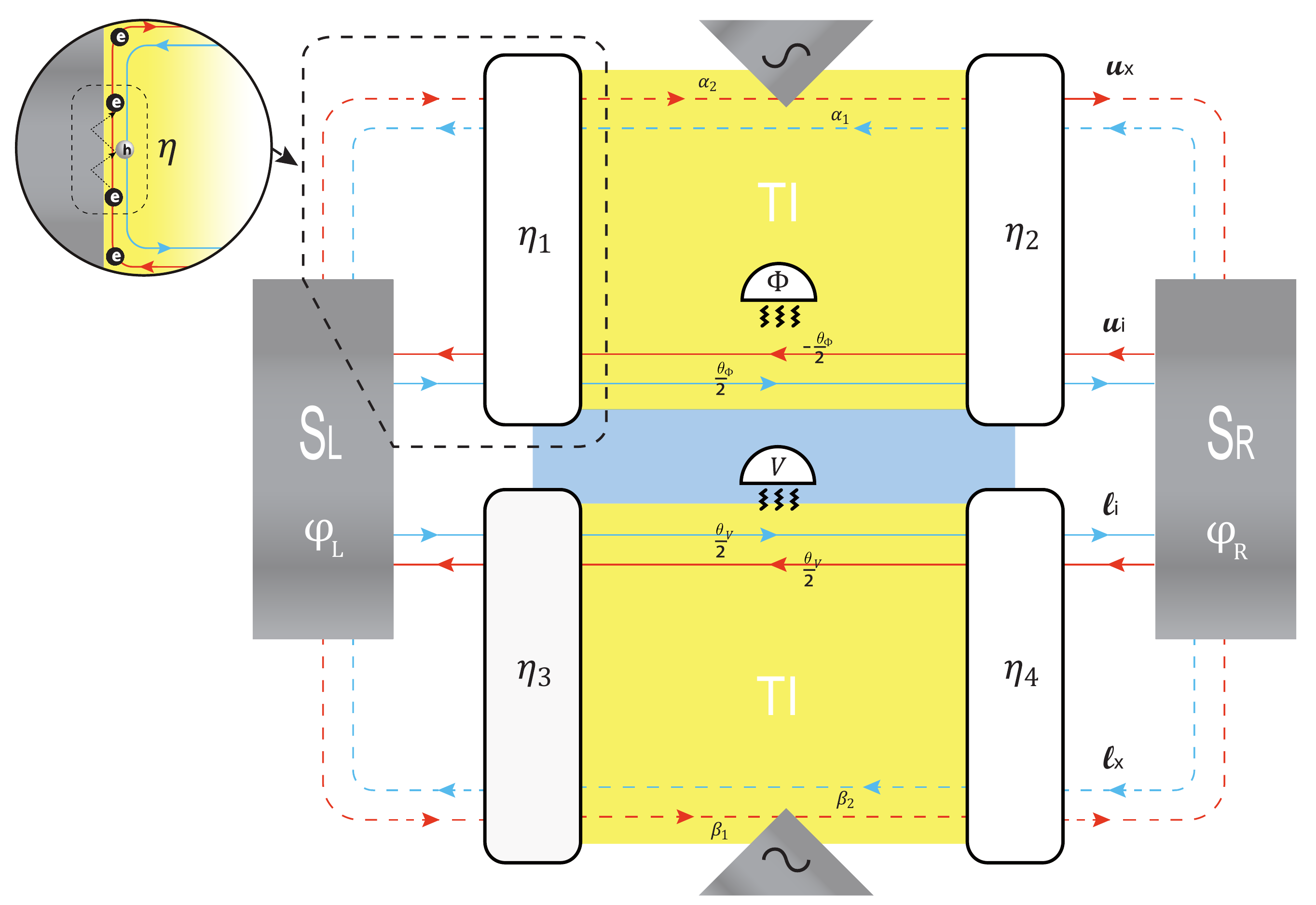}
    \caption{The full scheme of the system is depicted here by unfolding the 3D model of Fig.~\ref{3D}(b) in the main text, keeping fixed the frontal side in the $z$-$x$-plane (cyan area) in between the two superconductors ($S_L$ and $S_R$ in gray), and tilting the upper TI-plane along the $z$-axis and the lower TI-plane in the opposite direction, such that the top of the first TI-plane and the bottom of the second one are coplanar. In this scheme, the edges in the $z$-$x$-plane (frontal side) are defined as internal and labeled with $u_i$ and $\ell_i$ (blue/red solid lines) instead the edge modes running in the backside are referred to as external channels (blue/red dashed lines), and labeled with $u_x$ and $\ell_x$ for the upper and lower TI-plane respectively.  In the model the coherence of the branches is suppressed assuming to insert dephasing angles $\alpha_1,\alpha_2,\beta_1,\beta_2\overset{i.i.d.}{\sim} U_{[0,2\pi]}$ and averaging over them. As in the main text, we considered the case in which the $V$ field is applied along the $\ell_i$ edge of the lower TI-plane and the $\Phi$ field is applied along the $u_i$ edge of the upper TI-plane. The white boxes at each interface between the TI-planes and the superconductors depict the beam splitters (BSs) which model the scattering of the particle among the internal and external channels of each TI respectively. Notice that, in case the beam splitters are purely transmitting, there is no coupling between the external and internal edges corresponding to the limit of $\eta=0$ we used in the main text for the analytical derivation of the CPR.
    In the inset an incoming electron, impinging toward the left SC, emerges an another electron with the same spin on the opposite counter-propagating edge of the same TI-plane, after an even number of perfect Andreev reflections.} 
    \label{full_scheme} 
\end{figure*}

Here we discuss in detail the model we use in the main text.
The full scheme of the system is sketched in Fig.~$\ref{full_scheme}$ (details in the caption). 
Following the arrangement of the local fields discussed in the main text, here we considered the application of the $V$-field on the internal edge of the lower TI-plane together with the application of the $\Phi$-field on the internal edge of the upper one [Fig.~$\ref{full_scheme}$] (later we will discuss how to go beyond to this semplification). The internal (external) edges corresponds to the frontside (backside) edges of setup shown in Fig.~\ref{3D}(b)
of the main text.
The model consists of four beam-splitters (BSs) which describe effectively the contact interfaces between the superconductors and the TI-planes. 
This is needed - also in case of ideal interfaces - in order to take into account those scattering processes of particles which involve both the (internal and external) edges of a same TI-plane by means of ordinary reflection processes.
Intriguingly those processes may be also mediated by multiple Andreev reflections. 
Indeed, for example, an incoming electron toward the SC, can emerges an another electron with the same spin on the opposite counter-propagating edge of the same TI-plane, after an even number of perfect Andreev reflections (see the inset in Fig.~$\ref{full_scheme}$) independently of their local or nonlocal nature. 
Anyway, we will see in a moment that if the edge modes running in the backside part of the device (hereinafter referred to as “external edges”, namely the dashed lines depicted in Fig.~$\ref{full_scheme}$) are long enough with respect to $\ell_{\phi}$ or properly dephased with a voltage probe their only action is only to suppress the critical current not affecting the general conclusions of the paper. 
We modeled this mechanism by introducing a loss parameter $\eta_j\in [0,1]$ with $j=1,2,3,4$; which describes the reflectance probability of the BS at each interface, such that:
\begin{equation}
\eta_j=\abs{r_j}^2=1-\abs{t_j}^2
\end{equation}

where $r_j$ and $t_j$ represent respectively the reflectance and transmittance amplitudes of the $j^{th}$ BS. In particular, if $\eta_j=0$ there are no losses, namely the BS is perfectly transmissive and no electrons are reflected from internal to external  modes of Fig.~$\ref{full_scheme}$. Conversely if $\eta_j=1$ the BS is perfectly reflective and all the electrons reaching the superconductive contact from the internal edges will be reflected onto the external modes (and vice versa), no Andreev reflection are possible in that case and Josephson current is null. For the sake of simplicity, in our calculation, we considered all the beam splitters to be characterized by the same reflectance amplitude, such that $\eta_j=\eta$ $\forall j=1,2,3,4$ but all the results given can be easily generalized to a less symmetric case.

Following the prescription presented in~\cite{beenakker1992}, in order to derive the Josephson current of the model, we have to calculate the Andreev bound state energies $\epsilon_p$, solving the following self-consistent secular problem:
\begin{equation}
Det\qty[e^{i\arccos{\qty(\epsilon_p/\Delta_0)}}\mathbb{1}-s_As_N]=0
\label{eq:SM_Andreev}
\end{equation}
Here $s_A$ and $s_N$ are respectively the Andreev scattering reflection matrix and the scattering matrix describing the weak-link in the short-junction limit with ideal interfaces; which in turn (according the notation in \cite{beenakker1992}) take the following form:
\begin{equation}
\label{def_sN_s0}
s_N=\mqty(s_0 & \O \\ \O  & s_0^*);\quad s_A=\mqty(\O  & r_A\\r_A^* & \O )
\end{equation}
In Eq.($\ref{def_sN_s0}$), the scattering matrix component $r_A^*$, which relates the electrons impinging the SCs to the respective Andreev-reflected holes, takes the following form:
\begin{widetext}
\begin{equation}
\label{rA_Definitive_Model}
\mqty(b_{uxL}^{{\color{red}\downarrow}}\\ b_{uiL}^{{\color{blue}\uparrow}} \\b_{liL}^{{\color{blue}\uparrow}}\\b_{lxL}^{{\color{red}\downarrow}}\\b_{uxR}^{{\color{blue}\uparrow}}\\b_{uiR}^{{\color{red}\downarrow}}\\b_{liR}^{{\color{red}\downarrow}}\\b_{lxR}^{{\color{blue}\uparrow}})_{in}=
\begin{pmatrix}
 \begin{pmatrix}
\abs{\Lambda_{Lx}} &  0& 0 & i\abs{X_{Lx}}\\ 
 0&  \abs{\Lambda_{Li}}& i\abs{X_{Li}} &0 \\ 
0 &  i\abs{X_{Li}} & \abs{\Lambda_{Li}} &0 \\ 
 i\abs{X_{Lx}}& 0 & 0 &\abs{\Lambda_{Lx}} 
\end{pmatrix}e^{i \phi_L}& \O \\ 
\O& \begin{pmatrix}
\abs{\Lambda_{Rx}} &  0& 0 &i\abs{ X_{Rx}}\\ 
 0&  \abs{\Lambda_{Ri}}& i\abs{X_{Ri}} &0 \\ 
0 &  i\abs{X_{Ri}} & \abs{\Lambda_{Ri}} &0 \\ 
 i\abs{X_{Rx}}& 0 & 0 &\abs{\Lambda_{Rx}} 
\end{pmatrix}e^{i \phi_R}
\end{pmatrix}_{r_A^*}
\mqty(c_{uxL}^{{\color{blue}\uparrow}}\\ c_{uiL}^{{\color{red}\downarrow}} \\c_{liL}^{{\color{red}\downarrow}}\\c_{lxL}^{{\color{blue}\uparrow}}\\c_{uxR}^{{\color{red}\downarrow}}\\c_{uiR}^{{\color{blue}\uparrow}}\\c_{liR}^{{\color{blue}\uparrow}}\\c_{lxR}^{{\color{red}\downarrow}})_{out}
\end{equation}
\end{widetext}
It differs from the Andreev matrix of the main text because of the explicit presence of the edge modes running in the backside part of the device which add new scattering channels to the final structure (namely the external edges depicted in Fig.~$\ref{full_scheme}$ and labeled with $u_x$ and $\ell_x$ respectively). 
For this reason we have to enlarge the set of the splitting parameters, i.~e.  $\{\Lambda_{Sn},X_{Sn}\}$ as presented in Eq.~($\ref{rA_Definitive_Model}$), accounting for the local and non-local splitting of Cooper pairs on each side of the junction $S=L,R$, and along the specific set of internal and external channels $n=i,x$.
One may have noticed, both from the scheme of Fig.~$\ref{full_scheme}$ and the structure itself of the scattering matrix of Eq.~($\ref{rA_Definitive_Model}$), that the splitting of CPs along the internal and external edges of the model are related to independent mechanisms, which are ruled by the set of constrain equations $|\Lambda_{Sn}|^2+|X_{Sn}|^2=1$ for $S=L,R$ and $n=i,x$, imposed on the relative strength of the local and non-local splitting amplitudes because of unitarity. 
At microscopical level this competitive role of LAR vs CAR processes at each interface depends on the strength of the Coulomb interaction between the edges.
Note also that the phase difference in the Josephson junction is defined as $\phi=\phi_L-\phi_R$.

The scattering matrix $s_N$, which describes the weak-link, does not couple electrons and holes, thus it takes a block-diagonal form in the electron-hole space as shown in Eq.~($\ref{def_sN_s0}$). Specifically, the block-matrix component $s_0$, which relates incoming and outgoing electrons only, assumes the following structure:
\begin{widetext}
\begin{equation}
\label{S0}
\mqty(c_{uxL}^{{\color{blue}\uparrow}}\\ c_{uiL}^{{\color{red}\downarrow}} \\c_{liL}^{{\color{red}\downarrow}}\\c_{lxL}^{{\color{blue}\uparrow}}\\c_{uxR}^{{\color{red}\downarrow}}\\c_{uiR}^{{\color{blue}\uparrow}}\\c_{liR}^{{\color{blue}\uparrow}}\\c_{lxR}^{{\color{red}\downarrow}})_{out}=
\begin{pmatrix}
\begin{matrix}
\begin{matrix}
0&A_2\\
A_1&0
\end{matrix}&\begin{matrix}
0&0\\
0&0
\end{matrix}\\
\begin{matrix}
0&0\\
0&0
\end{matrix}&\begin{matrix}
0&A_4\\
A_3&0
\end{matrix}
\end{matrix}&\begin{matrix}
\begin{matrix}
D_1&0\\
0&D_2
\end{matrix}&\begin{matrix}
0&0\\
0&0
\end{matrix}\\
\begin{matrix}
0&0\\
0&0
\end{matrix}&\begin{matrix}
D_3&0\\
0&D_4
\end{matrix}
\end{matrix}\\
\begin{matrix}
\begin{matrix}
C_1&0\\
0&C_2
\end{matrix}&\begin{matrix}
0&0\\
0&0
\end{matrix}\\
\begin{matrix}
0&0\\
0&0
\end{matrix}&\begin{matrix}
C_3&0\\
0&C_4
\end{matrix}
\end{matrix}&\begin{matrix}
\begin{matrix}
0&B_2\\
B_1&0
\end{matrix}&\begin{matrix}
0&0\\
0&0
\end{matrix}\\
\begin{matrix}
0&0\\
0&0
\end{matrix}&\begin{matrix}
0&B_4\\
B_3&0
\end{matrix}
\end{matrix}
\end{pmatrix}_{s_0}
\mqty(c_{uxL}^{{\color{red}\downarrow}}\\ c_{uiL}^{{\color{blue}\uparrow}} \\c_{liL}^{{\color{blue}\uparrow}}\\c_{lxL}^{{\color{red}\downarrow}}\\c_{uxR}^{{\color{blue}\uparrow}}\\c_{uiR}^{{\color{red}\downarrow}}\\c_{liR}^{{\color{red}\downarrow}}\\c_{lxR}^{{\color{blue}\uparrow}})_{in}
\end{equation}
In which
\begin{equation}
\label{Entries}
\begin{matrix}
A_1=r_1+\frac{t_1^2r_2e^{i\alpha_2}e^{-i\theta_{\rm \Phi}/2}}{1-r_1r_2e^{i\alpha_2}e^{-i\theta_{\rm \Phi}/2}};&A_2=r_1+\frac{t_1^2r_2e^{i\alpha_1}e^{i\theta_{\rm \Phi}/2}}{1-r_1r_2e^{i\alpha_1}e^{i\theta_{\rm \Phi}/2}};&A_3=r_3+\frac{t_3^2r_4e^{i\beta_2}e^{-i\theta_{V}/2}}{1-r_3r_4e^{i\beta_2}e^{-i\theta_{V}/2}};& A_4=r_3+\frac{t_3^2r_4e^{i\beta_1}e^{i\theta_{V}/2}}{1-r_3r_4e^{i\beta_1}e^{i\theta_{V}/2}}\\
&&&\\
B_1=r_1+\frac{t_2^2r_1e^{i\alpha_1}e^{i\theta_{\rm \Phi}/2}}{1-r_1r_2e^{i\alpha_1}e^{i\theta_{\rm \Phi}/2}};&B_2=r_1+\frac{t_2^2r_1e^{i\alpha_2}e^{-i\theta_{\rm \Phi}/2}}{1-r_1r_2e^{i\alpha_2}e^{-i\theta_{\rm \Phi}/2}};&B_3=r_4+\frac{t_4^2r_3e^{i\beta_1}e^{i\theta_{V}/2}}{1-r_3r_4e^{i\beta_1}e^{i\theta_{V}/2}};& B_4=r_4+\frac{t_4^2r_3e^{i\beta_2}e^{-i\theta_{V}/2}}{1-r_3r_4e^{i\beta_2}e^{-i\theta_{V}/2}}\\
&&&\\
C_1=\frac{t_1t_2e^{i\alpha_2}}{1-r_1r_2e^{i\alpha_2}e^{-i\theta_{\rm \Phi}/2}};&C_2=\frac{t_1t_2e^{i\theta_{\rm \Phi}/2}}{1-r_1r_2e^{i\alpha_1}e^{i\theta_{\rm \Phi}/2}};&C_3=\frac{t_3t_4e^{-i\theta_{V}/2}}{1-r_3r_4e^{i\beta_2}e^{-i\theta_{V}/2}};&C_4=\frac{t_3t_4e^{i\beta_1}}{1-r_3r_4e^{i\beta_2}e^{i\theta_{V}/2}}
&&&\\
D_1=\frac{t_1t_2e^{i\alpha_1}}{1-r_1r_2e^{i\alpha_1}e^{i\theta_{\rm \Phi}/2}};&D_2=\frac{t_1t_2e^{-i\theta_{\rm \Phi}/2}}{1-r_1r_2e^{i\alpha_2}e^{-i\theta_{\rm \Phi}/2}};&D_3=\frac{t_3t_4e^{i\theta_{V}/2}}{1-r_3r_4e^{i\beta_1}e^{i\theta_{V}/2}};&D_4=\frac{t_3t_4e^{i\beta_1}}{1-r_3r_4e^{i\beta_2}e^{-i\theta_{V}/2}}
\end{matrix}
\end{equation}
\end{widetext}
A similar relation links the incoming and outgoing holes through $s_0^*$. 
As an example of the derivation of the non-null entries of Eq.~($\ref{S0}$), let us explicit the calculation of the term $A_1$, which relates an incoming electron from the upper-external branch on the left side (labeled by $uxL$), with an outgoing electron with the same spin on the internal-upper edge, again at interface with $S_L$ (labeled by $uiL$): 
\begin{widetext}
\begin{align}
\label{A1}
c_{uxL}^{{\color{red}\downarrow}}\rightarrow c_{uiL}^{{\color{red}\downarrow}}:\nonumber\\&A_1=r_1+t_1e^{i\alpha_2}r_2e^{-i\theta_{\rm \Phi}/2}t_1+t_1e^{i\alpha_2}\vdot r_2e^{-i\theta_{\rm \Phi}/2} r_1e^{i\alpha_2}\vdot r_2e^{-i\theta_{\rm \Phi}/2}t_1+\dots\nonumber\\&=r_1+t_1^2r_2e^{i\alpha_2}e^{-i\theta_{\rm \Phi}/2}\sum_{n=0}^{\infty}(r_2r_1e^{i\alpha_2}e^{-i\theta_{\rm \Phi}/2})^n\nonumber\\&=r_1+\frac{t_1^2r_2e^{i\alpha_2}e^{-i\theta_{\rm \Phi}/2}}{1-r_1r_2e^{i\alpha_2}e^{-i\theta_{\rm \Phi}/2}}
\end{align}
\end{widetext}
Multiple reflections between the different BSs have been taken into account, as results from the geometrical series in Eq.~($\ref{A1}$). In the previous equation we also introduce the phases acquired during the 
evolution along the external edges labeled as $\alpha_i$ and $\beta_i$. Those phases are introduced in order to effectively describe the dephasing processes since, in the end, we average the physical quantities over them (namely $\alpha_i,\beta_i\overset{i.i.d.}{\sim} U_{[0,2\pi]}$).
So by the set of the previous equations we can calculate the Josephson current $J(\phi,\theta_{\rm \Phi},\theta_{V},\alpha_1,\alpha_2,\beta_1,\beta_2)$ where the dependence over $T$ and $\eta$ is implicitly assumed. The final value of this quantity, in our results is obtained by the mentioned averaging procedure, i.~e. $\bar{J}(\phi,\theta_{\rm \Phi},\theta_{V})=\frac{1}{(2\pi)^4}\int_0^{2\pi}d\alpha_1d\alpha_2d\beta_1d\beta_2~J(\phi,\theta_{\rm \Phi},\theta_{V},\alpha_1,\alpha_2,\beta_1,\beta_2)$.
In the main text we show the numerical results for the Josephson current at finite $\eta$ and temperature $T$. In section~\ref{single_shot_regime} we analytically derive, with the same method, the Josephson current as perturbative expansion in $(1-\eta)$.

\section{Configuration of the fields}\label{Appendix_D}

The most general scheme of the application of the local fields along the edge states of the system is depicted in Fig.~$\ref{Fields_general_scheme}$. We represents four local fields contributions divided among the different edge states such that each set of helical modes (belonging to the upper and lower TI-plane respectively) is interested by one $V$-type field (which describing the TRS terms) and one $\Phi$-type field (describing TRS breaking terms). Here we labelled each manipulation angle, associated with the corresponding term, $\theta_{Vn},\theta_{\Phi n}$ with the index $n=u,\ell$ to indicate the pertinent upper or lower edge (plane) of application.

Within this picture, by following the same procedure employed in the main text for the calculation of the Josephson current, we obtained - in case of no losses ($\eta=0$) - the following result:
\begin{widetext}
\begin{equation}
\label{J_current_analytical_B_Bs}
J(\phi)= 4\frac{e\Delta_0}{\hbar}\sum_{\nu=\pm}\Bigg\{\sin{\left(\frac{\bar{\theta}_{\rm \Phi}+\phi}{2} +\nu \tan^{-1}{\left(\sqrt{\frac{1-\Gamma_{}}{1+\Gamma_{}}}\right)}\right)}\tanh{\qty[\frac{\Delta_0}{2k_BT}\cos{\left(\frac{\bar{\theta}_{\rm \Phi}+\phi}{2} +\nu\tan^{-1}{\left(\sqrt{\frac{1-\Gamma_{}}{1+\Gamma_{}}}\right)}\right)}]}\Bigg\} 
\end{equation}
\end{widetext}
which appears with the same functional form already presented in the main text. In particular, $\bar{\theta}_{\rm \Phi}=\frac{\theta_{\rm \Phi \textit{u}}+\theta_{\rm \Phi \ell}}{2}$ is given by the sum of the two separated $\Phi$ contributions. The quantity $\bar{\theta}_{\rm \Phi}$ affect the CPR with a global phase shifting and one immediately see that for $\bar{\theta}_{\Phi}\neq 0$ one could find an anomalous current, i.~e. Josephson current at $\phi=0$. This is consistent with the fact that in general anomalous current can be generated by the breaking of TRS. Anyway if the TRS is broken, at local level, but in an exactly opposite way, such as $\theta_{\Phi \textit{u}}=-\theta_{\Phi \ell}$, the anomalous current disappear since $\bar{\theta}_{\rm \Phi}=0$.
At the same time also the function $\Gamma$ is generalized:
\begin{figure*}[ht]
\centering
    \includegraphics[width=.6\textwidth]{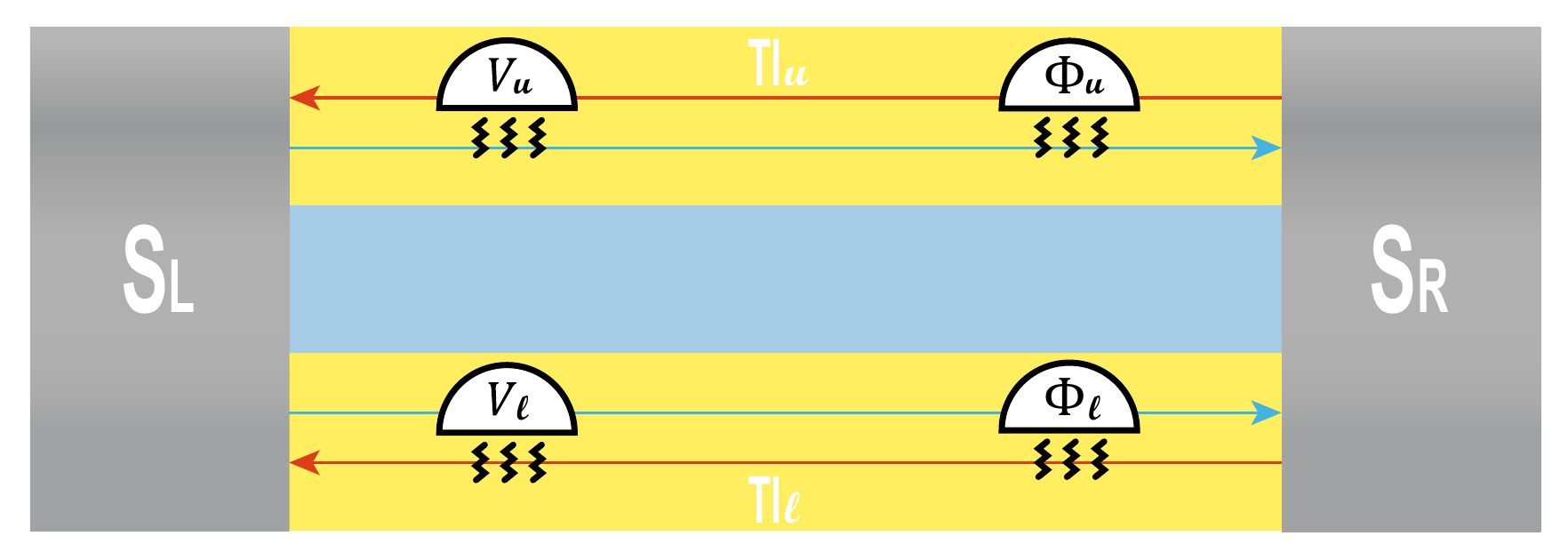}
    \caption{General application of the fields along the edges states of the system.} 
    \label{Fields_general_scheme} 
\end{figure*}
\begin{align}
\label{Gamma}
\Gamma=\cos{\left(\frac{\Delta \theta_{\rm V}}{2}\right)}\abs{X_L}\abs{X_R}+\cos{\left(\frac{\Delta\theta_{\rm \Phi}}{2}\right)}\abs{\Lambda_L}\abs{\Lambda_R},
\end{align}
where we note that the effective action of $\theta_{Vn}$ and $\theta_{\Phi n}$
is given by their differential mode $\Delta \theta_{\rm V}=\theta_{\rm V \textit{u}}-\theta_{\rm V \ell}$ and $\Delta\theta_{V}=\theta_{\rm \Phi \textit{u}}-\theta_{\rm \Phi \ell}$. So only the difference between the local action in the upper and lower edges of both the $V$ and $\Phi$ terms effectively contributes on the modification of the shape of the CPR. Intriguingly, the approaches suggested to generate the fields operate (by construction) on the \emph{differential} mode which are the required terms which modify the $\Gamma$ of Eq.~(\ref{Gamma}).
In this way we still preserve the selective action of the two different fields which operate in a targeted manner on the local and non-local components of the current as discussed in the main text. 

\bibliographystyle{apsrev4-1}
\bibliography{biblio}

\begin{thebibliography}{62}%
\makeatletter
\providecommand \@ifxundefined [1]{%
 \@ifx{#1\undefined}
}%
\providecommand \@ifnum [1]{%
 \ifnum #1\expandafter \@firstoftwo
 \else \expandafter \@secondoftwo
 \fi
}%
\providecommand \@ifx [1]{%
 \ifx #1\expandafter \@firstoftwo
 \else \expandafter \@secondoftwo
 \fi
}%
\providecommand \natexlab [1]{#1}%
\providecommand \enquote  [1]{``#1''}%
\providecommand \bibnamefont  [1]{#1}%
\providecommand \bibfnamefont [1]{#1}%
\providecommand \citenamefont [1]{#1}%
\providecommand \href@noop [0]{\@secondoftwo}%
\providecommand \href [0]{\begingroup \@sanitize@url \@href}%
\providecommand \@href[1]{\@@startlink{#1}\@@href}%
\providecommand \@@href[1]{\endgroup#1\@@endlink}%
\providecommand \@sanitize@url [0]{\catcode `\\12\catcode `\$12\catcode
  `\&12\catcode `\#12\catcode `\^12\catcode `\_12\catcode `\%12\relax}%
\providecommand \@@startlink[1]{}%
\providecommand \@@endlink[0]{}%
\providecommand \url  [0]{\begingroup\@sanitize@url \@url }%
\providecommand \@url [1]{\endgroup\@href {#1}{\urlprefix }}%
\providecommand \urlprefix  [0]{URL }%
\providecommand \Eprint [0]{\href }%
\providecommand \doibase [0]{http://dx.doi.org/}%
\providecommand \selectlanguage [0]{\@gobble}%
\providecommand \bibinfo  [0]{\@secondoftwo}%
\providecommand \bibfield  [0]{\@secondoftwo}%
\providecommand \translation [1]{[#1]}%
\providecommand \BibitemOpen [0]{}%
\providecommand \bibitemStop [0]{}%
\providecommand \bibitemNoStop [0]{.\EOS\space}%
\providecommand \EOS [0]{\spacefactor3000\relax}%
\providecommand \BibitemShut  [1]{\csname bibitem#1\endcsname}%
\let\auto@bib@innerbib\@empty
\bibitem [{\citenamefont {Einstein}\ \emph {et~al.}(1935)\citenamefont
  {Einstein}, \citenamefont {Podolsky},\ and\ \citenamefont
  {Rosen}}]{Einstein1935}%
  \BibitemOpen
  \bibfield  {author} {\bibinfo {author} {\bibfnamefont {A.}~\bibnamefont
  {Einstein}}, \bibinfo {author} {\bibfnamefont {B.}~\bibnamefont {Podolsky}},
  \ and\ \bibinfo {author} {\bibfnamefont {N.}~\bibnamefont {Rosen}},\ }\href
  {\doibase 10.1103/PhysRev.47.777} {\bibfield  {journal} {\bibinfo  {journal}
  {Phys. Rev.}\ }\textbf {\bibinfo {volume} {47}},\ \bibinfo {pages} {777}
  (\bibinfo {year} {1935})}\BibitemShut {NoStop}%
\bibitem [{\citenamefont {Bell}(1966)}]{Bell1966}%
  \BibitemOpen
  \bibfield  {author} {\bibinfo {author} {\bibfnamefont {J.~S.}\ \bibnamefont
  {Bell}},\ }\href {\doibase 10.1103/RevModPhys.38.447} {\bibfield  {journal}
  {\bibinfo  {journal} {Rev. Mod. Phys.}\ }\textbf {\bibinfo {volume} {38}},\
  \bibinfo {pages} {447} (\bibinfo {year} {1966})}\BibitemShut {NoStop}%
\bibitem [{\citenamefont {Clauser}\ \emph {et~al.}(1969)\citenamefont
  {Clauser}, \citenamefont {Horne}, \citenamefont {Shimony},\ and\
  \citenamefont {Holt}}]{Clauser1969}%
  \BibitemOpen
  \bibfield  {author} {\bibinfo {author} {\bibfnamefont {J.~F.}\ \bibnamefont
  {Clauser}}, \bibinfo {author} {\bibfnamefont {M.~A.}\ \bibnamefont {Horne}},
  \bibinfo {author} {\bibfnamefont {A.}~\bibnamefont {Shimony}}, \ and\
  \bibinfo {author} {\bibfnamefont {R.~A.}\ \bibnamefont {Holt}},\ }\href
  {\doibase 10.1103/PhysRevLett.23.880} {\bibfield  {journal} {\bibinfo
  {journal} {Phys. Rev. Lett.}\ }\textbf {\bibinfo {volume} {23}},\ \bibinfo
  {pages} {880} (\bibinfo {year} {1969})}\BibitemShut {NoStop}%
\bibitem [{\citenamefont {Aspect}\ \emph {et~al.}(1981)\citenamefont {Aspect},
  \citenamefont {Grangier},\ and\ \citenamefont {Roger}}]{Aspect1981}%
  \BibitemOpen
  \bibfield  {author} {\bibinfo {author} {\bibfnamefont {A.}~\bibnamefont
  {Aspect}}, \bibinfo {author} {\bibfnamefont {P.}~\bibnamefont {Grangier}}, \
  and\ \bibinfo {author} {\bibfnamefont {G.}~\bibnamefont {Roger}},\ }\href
  {\doibase 10.1103/PhysRevLett.47.460} {\bibfield  {journal} {\bibinfo
  {journal} {Phys. Rev. Lett.}\ }\textbf {\bibinfo {volume} {47}},\ \bibinfo
  {pages} {460} (\bibinfo {year} {1981})}\BibitemShut {NoStop}%
\bibitem [{\citenamefont {Tittel}\ \emph {et~al.}(1998)\citenamefont {Tittel},
  \citenamefont {Brendel}, \citenamefont {Zbinden},\ and\ \citenamefont
  {Gisin}}]{Tittel1998}%
  \BibitemOpen
  \bibfield  {author} {\bibinfo {author} {\bibfnamefont {W.}~\bibnamefont
  {Tittel}}, \bibinfo {author} {\bibfnamefont {J.}~\bibnamefont {Brendel}},
  \bibinfo {author} {\bibfnamefont {H.}~\bibnamefont {Zbinden}}, \ and\
  \bibinfo {author} {\bibfnamefont {N.}~\bibnamefont {Gisin}},\ }\href
  {\doibase 10.1103/PhysRevLett.81.3563} {\bibfield  {journal} {\bibinfo
  {journal} {Phys. Rev. Lett.}\ }\textbf {\bibinfo {volume} {81}},\ \bibinfo
  {pages} {3563} (\bibinfo {year} {1998})}\BibitemShut {NoStop}%
\bibitem [{\citenamefont {Ekert}(1991)}]{Ekert1991}%
  \BibitemOpen
  \bibfield  {author} {\bibinfo {author} {\bibfnamefont {A.~K.}\ \bibnamefont
  {Ekert}},\ }\href {\doibase 10.1103/PhysRevLett.67.661} {\bibfield  {journal}
  {\bibinfo  {journal} {Phys. Rev. Lett.}\ }\textbf {\bibinfo {volume} {67}},\
  \bibinfo {pages} {661} (\bibinfo {year} {1991})}\BibitemShut {NoStop}%
\bibitem [{\citenamefont {Gisin}\ \emph {et~al.}(2002)\citenamefont {Gisin},
  \citenamefont {Ribordy}, \citenamefont {Tittel},\ and\ \citenamefont
  {Zbinden}}]{Gisin2002}%
  \BibitemOpen
  \bibfield  {author} {\bibinfo {author} {\bibfnamefont {N.}~\bibnamefont
  {Gisin}}, \bibinfo {author} {\bibfnamefont {G.}~\bibnamefont {Ribordy}},
  \bibinfo {author} {\bibfnamefont {W.}~\bibnamefont {Tittel}}, \ and\ \bibinfo
  {author} {\bibfnamefont {H.}~\bibnamefont {Zbinden}},\ }\href {\doibase
  10.1103/RevModPhys.74.145} {\bibfield  {journal} {\bibinfo  {journal} {Rev.
  Mod. Phys.}\ }\textbf {\bibinfo {volume} {74}},\ \bibinfo {pages} {145}
  (\bibinfo {year} {2002})}\BibitemShut {NoStop}%
\bibitem [{\citenamefont {Loss}\ and\ \citenamefont
  {DiVincenzo}(1998)}]{Loss1998}%
  \BibitemOpen
  \bibfield  {author} {\bibinfo {author} {\bibfnamefont {D.}~\bibnamefont
  {Loss}}\ and\ \bibinfo {author} {\bibfnamefont {D.~P.}\ \bibnamefont
  {DiVincenzo}},\ }\href {\doibase 10.1103/PhysRevA.57.120} {\bibfield
  {journal} {\bibinfo  {journal} {Phys. Rev. A}\ }\textbf {\bibinfo {volume}
  {57}},\ \bibinfo {pages} {120} (\bibinfo {year} {1998})}\BibitemShut
  {NoStop}%
\bibitem [{\citenamefont {Lorenzo}\ and\ \citenamefont
  {Nazarov}(2005)}]{Nazarov2005}%
  \BibitemOpen
  \bibfield  {author} {\bibinfo {author} {\bibfnamefont {A.~D.}\ \bibnamefont
  {Lorenzo}}\ and\ \bibinfo {author} {\bibfnamefont {Y.~V.}\ \bibnamefont
  {Nazarov}},\ }\href {\doibase 10.1103/PhysRevLett.94.210601} {\bibfield
  {journal} {\bibinfo  {journal} {Phys. Rev. Lett.}\ }\textbf {\bibinfo
  {volume} {94}},\ \bibinfo {pages} {210601} (\bibinfo {year}
  {2005})}\BibitemShut {NoStop}%
\bibitem [{\citenamefont {Burkard}\ \emph {et~al.}(2000)\citenamefont
  {Burkard}, \citenamefont {Loss},\ and\ \citenamefont
  {Sukhorukov}}]{Burkard2000}%
  \BibitemOpen
  \bibfield  {author} {\bibinfo {author} {\bibfnamefont {G.}~\bibnamefont
  {Burkard}}, \bibinfo {author} {\bibfnamefont {D.}~\bibnamefont {Loss}}, \
  and\ \bibinfo {author} {\bibfnamefont {E.~V.}\ \bibnamefont {Sukhorukov}},\
  }\href {\doibase 10.1103/PhysRevB.61.R16303} {\bibfield  {journal} {\bibinfo
  {journal} {Phys. Rev. B}\ }\textbf {\bibinfo {volume} {61}},\ \bibinfo
  {pages} {R16303} (\bibinfo {year} {2000})}\BibitemShut {NoStop}%
\bibitem [{\citenamefont {Recher}\ \emph {et~al.}(2001)\citenamefont {Recher},
  \citenamefont {Sukhorukov},\ and\ \citenamefont {Loss}}]{Recher2001}%
  \BibitemOpen
  \bibfield  {author} {\bibinfo {author} {\bibfnamefont {P.}~\bibnamefont
  {Recher}}, \bibinfo {author} {\bibfnamefont {E.~V.}\ \bibnamefont
  {Sukhorukov}}, \ and\ \bibinfo {author} {\bibfnamefont {D.}~\bibnamefont
  {Loss}},\ }\href {\doibase 10.1103/PhysRevB.63.165314} {\bibfield  {journal}
  {\bibinfo  {journal} {Phys. Rev. B}\ }\textbf {\bibinfo {volume} {63}},\
  \bibinfo {pages} {165314} (\bibinfo {year} {2001})}\BibitemShut {NoStop}%
\bibitem [{\citenamefont {Bena}\ \emph {et~al.}(2002)\citenamefont {Bena},
  \citenamefont {Vishveshwara}, \citenamefont {Balents},\ and\ \citenamefont
  {Fisher}}]{Bena2002}%
  \BibitemOpen
  \bibfield  {author} {\bibinfo {author} {\bibfnamefont {C.}~\bibnamefont
  {Bena}}, \bibinfo {author} {\bibfnamefont {S.}~\bibnamefont {Vishveshwara}},
  \bibinfo {author} {\bibfnamefont {L.}~\bibnamefont {Balents}}, \ and\
  \bibinfo {author} {\bibfnamefont {M.~P.~A.}\ \bibnamefont {Fisher}},\ }\href
  {\doibase 10.1103/PhysRevLett.89.037901} {\bibfield  {journal} {\bibinfo
  {journal} {Phys. Rev. Lett.}\ }\textbf {\bibinfo {volume} {89}},\ \bibinfo
  {pages} {037901} (\bibinfo {year} {2002})}\BibitemShut {NoStop}%
\bibitem [{\citenamefont {Kim}\ \emph {et~al.}(2004)\citenamefont {Kim},
  \citenamefont {Vishveshwara},\ and\ \citenamefont {Fradkin}}]{Kim2004}%
  \BibitemOpen
  \bibfield  {author} {\bibinfo {author} {\bibfnamefont {E.-A.}\ \bibnamefont
  {Kim}}, \bibinfo {author} {\bibfnamefont {S.}~\bibnamefont {Vishveshwara}}, \
  and\ \bibinfo {author} {\bibfnamefont {E.}~\bibnamefont {Fradkin}},\ }\href
  {\doibase 10.1103/PhysRevLett.93.266803} {\bibfield  {journal} {\bibinfo
  {journal} {Phys. Rev. Lett.}\ }\textbf {\bibinfo {volume} {93}},\ \bibinfo
  {pages} {266803} (\bibinfo {year} {2004})}\BibitemShut {NoStop}%
\bibitem [{\citenamefont {Hussein}\ \emph {et~al.}(2017)\citenamefont
  {Hussein}, \citenamefont {Braggio},\ and\ \citenamefont
  {Governale}}]{hussein2017}%
  \BibitemOpen
  \bibfield  {author} {\bibinfo {author} {\bibfnamefont {R.}~\bibnamefont
  {Hussein}}, \bibinfo {author} {\bibfnamefont {A.}~\bibnamefont {Braggio}}, \
  and\ \bibinfo {author} {\bibfnamefont {M.}~\bibnamefont {Governale}},\ }\href
  {https://doi.org/10.1002/pssb.201600603} {\bibfield  {journal} {\bibinfo
  {journal} {physica status solidi (b)}\ }\textbf {\bibinfo {volume} {254}},\
  \bibinfo {pages} {1600603} (\bibinfo {year} {2017})}\BibitemShut {NoStop}%
\bibitem [{\citenamefont {Hussein}\ \emph {et~al.}(2016)\citenamefont
  {Hussein}, \citenamefont {Jaurigue}, \citenamefont {Governale},\ and\
  \citenamefont {Braggio}}]{hussein2016}%
  \BibitemOpen
  \bibfield  {author} {\bibinfo {author} {\bibfnamefont {R.}~\bibnamefont
  {Hussein}}, \bibinfo {author} {\bibfnamefont {L.}~\bibnamefont {Jaurigue}},
  \bibinfo {author} {\bibfnamefont {M.}~\bibnamefont {Governale}}, \ and\
  \bibinfo {author} {\bibfnamefont {A.}~\bibnamefont {Braggio}},\ }\href
  {https://doi.org/10.1103/PhysRevB.94.235134} {\bibfield  {journal} {\bibinfo
  {journal} {Physical Review B}\ }\textbf {\bibinfo {volume} {94}},\ \bibinfo
  {pages} {235134} (\bibinfo {year} {2016})}\BibitemShut {NoStop}%
\bibitem [{\citenamefont {Sato}\ \emph {et~al.}(2010)\citenamefont {Sato},
  \citenamefont {Loss},\ and\ \citenamefont {Tserkovnyak}}]{Sato2010}%
  \BibitemOpen
  \bibfield  {author} {\bibinfo {author} {\bibfnamefont {K.}~\bibnamefont
  {Sato}}, \bibinfo {author} {\bibfnamefont {D.}~\bibnamefont {Loss}}, \ and\
  \bibinfo {author} {\bibfnamefont {Y.}~\bibnamefont {Tserkovnyak}},\ }\href
  {\doibase 10.1103/PhysRevLett.105.226401} {\bibfield  {journal} {\bibinfo
  {journal} {Phys. Rev. Lett.}\ }\textbf {\bibinfo {volume} {105}},\ \bibinfo
  {pages} {226401} (\bibinfo {year} {2010})}\BibitemShut {NoStop}%
\bibitem [{\citenamefont {Chen}\ and\ \citenamefont {Li}(2012)}]{Chen2012}%
  \BibitemOpen
  \bibfield  {author} {\bibinfo {author} {\bibfnamefont {Y.-X.}\ \bibnamefont
  {Chen}}\ and\ \bibinfo {author} {\bibfnamefont {S.-W.}\ \bibnamefont {Li}},\
  }\href {\doibase 10.1209/0295-5075/97/40003} {\bibfield  {journal} {\bibinfo
  {journal} {Europhys. Lett.}\ }\textbf {\bibinfo {volume} {97}},\ \bibinfo
  {pages} {40003} (\bibinfo {year} {2012})}\BibitemShut {NoStop}%
\bibitem [{\citenamefont {Virtanen}\ and\ \citenamefont
  {Recher}(2012)}]{Virtanen2012}%
  \BibitemOpen
  \bibfield  {author} {\bibinfo {author} {\bibfnamefont {P.}~\bibnamefont
  {Virtanen}}\ and\ \bibinfo {author} {\bibfnamefont {P.}~\bibnamefont
  {Recher}},\ }\href {\doibase 10.1103/PhysRevB.85.035310} {\bibfield
  {journal} {\bibinfo  {journal} {Phys. Rev. B}\ }\textbf {\bibinfo {volume}
  {85}},\ \bibinfo {pages} {035310} (\bibinfo {year} {2012})}\BibitemShut
  {NoStop}%
\bibitem [{\citenamefont {Reinthaler}\ \emph {et~al.}(2013)\citenamefont
  {Reinthaler}, \citenamefont {Recher},\ and\ \citenamefont
  {Hankiewicz}}]{Reinthaler2013}%
  \BibitemOpen
  \bibfield  {author} {\bibinfo {author} {\bibfnamefont {R.~W.}\ \bibnamefont
  {Reinthaler}}, \bibinfo {author} {\bibfnamefont {P.}~\bibnamefont {Recher}},
  \ and\ \bibinfo {author} {\bibfnamefont {E.~M.}\ \bibnamefont {Hankiewicz}},\
  }\href {\doibase 10.1103/PhysRevLett.110.226802} {\bibfield  {journal}
  {\bibinfo  {journal} {Phys. Rev. Lett.}\ }\textbf {\bibinfo {volume} {110}},\
  \bibinfo {pages} {226802} (\bibinfo {year} {2013})}\BibitemShut {NoStop}%
\bibitem [{\citenamefont {Choi}(2014)}]{Choi2014}%
  \BibitemOpen
  \bibfield  {author} {\bibinfo {author} {\bibfnamefont {M.-S.}\ \bibnamefont
  {Choi}},\ }\href {\doibase 10.1103/PhysRevB.89.045137} {\bibfield  {journal}
  {\bibinfo  {journal} {Phys. Rev. B}\ }\textbf {\bibinfo {volume} {89}},\
  \bibinfo {pages} {045137} (\bibinfo {year} {2014})}\BibitemShut {NoStop}%
\bibitem [{\citenamefont {Sato}\ and\ \citenamefont
  {Tserkovnyak}(2014)}]{Sato2014}%
  \BibitemOpen
  \bibfield  {author} {\bibinfo {author} {\bibfnamefont {K.}~\bibnamefont
  {Sato}}\ and\ \bibinfo {author} {\bibfnamefont {Y.}~\bibnamefont
  {Tserkovnyak}},\ }\href {\doibase 10.1103/PhysRevB.90.045419} {\bibfield
  {journal} {\bibinfo  {journal} {Phys. Rev. B}\ }\textbf {\bibinfo {volume}
  {90}},\ \bibinfo {pages} {045419} (\bibinfo {year} {2014})}\BibitemShut
  {NoStop}%
\bibitem [{\citenamefont {Cr{\'e}pin}\ and\ \citenamefont
  {Trauzettel}(2014)}]{crepin2014}%
  \BibitemOpen
  \bibfield  {author} {\bibinfo {author} {\bibfnamefont {F.}~\bibnamefont
  {Cr{\'e}pin}}\ and\ \bibinfo {author} {\bibfnamefont {B.}~\bibnamefont
  {Trauzettel}},\ }\href
  {https://journals.aps.org/prl/abstract/10.1103/PhysRevLett.112.077002}
  {\bibfield  {journal} {\bibinfo  {journal} {Phys. Rev. lett.}\ }\textbf
  {\bibinfo {volume} {112}},\ \bibinfo {pages} {077002} (\bibinfo {year}
  {2014})}\BibitemShut {NoStop}%
\bibitem [{\citenamefont {Veldhorst}\ \emph {et~al.}(2014)\citenamefont
  {Veldhorst}, \citenamefont {Hoek}, \citenamefont {Snelder}, \citenamefont
  {Hilgenkamp}, \citenamefont {Golubov},\ and\ \citenamefont
  {Brinkman}}]{Veldhorst2014}%
  \BibitemOpen
  \bibfield  {author} {\bibinfo {author} {\bibfnamefont {M.}~\bibnamefont
  {Veldhorst}}, \bibinfo {author} {\bibfnamefont {M.}~\bibnamefont {Hoek}},
  \bibinfo {author} {\bibfnamefont {M.}~\bibnamefont {Snelder}}, \bibinfo
  {author} {\bibfnamefont {H.}~\bibnamefont {Hilgenkamp}}, \bibinfo {author}
  {\bibfnamefont {A.~A.}\ \bibnamefont {Golubov}}, \ and\ \bibinfo {author}
  {\bibfnamefont {A.}~\bibnamefont {Brinkman}},\ }\href {\doibase
  10.1103/PhysRevB.90.035428} {\bibfield  {journal} {\bibinfo  {journal} {Phys.
  Rev. B}\ }\textbf {\bibinfo {volume} {90}},\ \bibinfo {pages} {035428}
  (\bibinfo {year} {2014})}\BibitemShut {NoStop}%
\bibitem [{\citenamefont {Zhang}\ \emph {et~al.}(2015)\citenamefont {Zhang},
  \citenamefont {Deng}, \citenamefont {Sun},\ and\ \citenamefont
  {Qiao}}]{Zhang2015}%
  \BibitemOpen
  \bibfield  {author} {\bibinfo {author} {\bibfnamefont {Y.-T.}\ \bibnamefont
  {Zhang}}, \bibinfo {author} {\bibfnamefont {X.}~\bibnamefont {Deng}},
  \bibinfo {author} {\bibfnamefont {Q.-F.}\ \bibnamefont {Sun}}, \ and\
  \bibinfo {author} {\bibfnamefont {Z.}~\bibnamefont {Qiao}},\ }\href
  {http://dx.doi.org/10.1038/srep14892} {\bibfield  {journal} {\bibinfo
  {journal} {Sci. Rep.}\ }\textbf {\bibinfo {volume} {5}},\ \bibinfo {pages}
  {14892} (\bibinfo {year} {2015})},\ \bibinfo {note} {article}\BibitemShut
  {NoStop}%
\bibitem [{\citenamefont {Str\"om}\ \emph {et~al.}(2015)\citenamefont
  {Str\"om}, \citenamefont {Johannesson},\ and\ \citenamefont
  {Recher}}]{Strom2015}%
  \BibitemOpen
  \bibfield  {author} {\bibinfo {author} {\bibfnamefont {A.}~\bibnamefont
  {Str\"om}}, \bibinfo {author} {\bibfnamefont {H.}~\bibnamefont
  {Johannesson}}, \ and\ \bibinfo {author} {\bibfnamefont {P.}~\bibnamefont
  {Recher}},\ }\href {\doibase 10.1103/PhysRevB.91.245406} {\bibfield
  {journal} {\bibinfo  {journal} {Phys. Rev. B}\ }\textbf {\bibinfo {volume}
  {91}},\ \bibinfo {pages} {245406} (\bibinfo {year} {2015})}\BibitemShut
  {NoStop}%
\bibitem [{\citenamefont {Wang}\ \emph {et~al.}(2015)\citenamefont {Wang},
  \citenamefont {Hao},\ and\ \citenamefont {Chan}}]{Wang2015}%
  \BibitemOpen
  \bibfield  {author} {\bibinfo {author} {\bibfnamefont {J.}~\bibnamefont
  {Wang}}, \bibinfo {author} {\bibfnamefont {L.}~\bibnamefont {Hao}}, \ and\
  \bibinfo {author} {\bibfnamefont {K.~S.}\ \bibnamefont {Chan}},\ }\href
  {\doibase 10.1103/PhysRevB.91.085415} {\bibfield  {journal} {\bibinfo
  {journal} {Phys. Rev. B}\ }\textbf {\bibinfo {volume} {91}},\ \bibinfo
  {pages} {085415} (\bibinfo {year} {2015})}\BibitemShut {NoStop}%
\bibitem [{\citenamefont {Hou}\ \emph {et~al.}(2016)\citenamefont {Hou},
  \citenamefont {Xing}, \citenamefont {Guo},\ and\ \citenamefont
  {Sun}}]{Hou2016}%
  \BibitemOpen
  \bibfield  {author} {\bibinfo {author} {\bibfnamefont {Z.}~\bibnamefont
  {Hou}}, \bibinfo {author} {\bibfnamefont {Y.}~\bibnamefont {Xing}}, \bibinfo
  {author} {\bibfnamefont {A.-M.}\ \bibnamefont {Guo}}, \ and\ \bibinfo
  {author} {\bibfnamefont {Q.-F.}\ \bibnamefont {Sun}},\ }\href {\doibase
  10.1103/PhysRevB.94.064516} {\bibfield  {journal} {\bibinfo  {journal} {Phys.
  Rev. B}\ }\textbf {\bibinfo {volume} {94}},\ \bibinfo {pages} {064516}
  (\bibinfo {year} {2016})}\BibitemShut {NoStop}%
\bibitem [{\citenamefont {Islam}\ \emph {et~al.}(2017)\citenamefont {Islam},
  \citenamefont {Dutta},\ and\ \citenamefont {Saha}}]{Islam2017}%
  \BibitemOpen
  \bibfield  {author} {\bibinfo {author} {\bibfnamefont {S.~F.}\ \bibnamefont
  {Islam}}, \bibinfo {author} {\bibfnamefont {P.}~\bibnamefont {Dutta}}, \ and\
  \bibinfo {author} {\bibfnamefont {A.}~\bibnamefont {Saha}},\ }\href {\doibase
  10.1103/PhysRevB.96.155429} {\bibfield  {journal} {\bibinfo  {journal} {Phys.
  Rev. B}\ }\textbf {\bibinfo {volume} {96}},\ \bibinfo {pages} {155429}
  (\bibinfo {year} {2017})}\BibitemShut {NoStop}%
\bibitem [{\citenamefont {B.~Andrei~Bernevig}(2006)}]{Bernevig2006}%
  \BibitemOpen
  \bibfield  {author} {\bibinfo {author} {\bibfnamefont {S.-C.~Z.}\
  \bibnamefont {B.~Andrei~Bernevig}, \bibfnamefont {Taylor L.~Hughes}},\ }\href
  {https://doi.org/10.1126/science.1133734} {\bibfield  {journal} {\bibinfo
  {journal} {Science}\ }\textbf {\bibinfo {volume} {314}} (\bibinfo {year}
  {2006})}\BibitemShut {NoStop}%
\bibitem [{\citenamefont {Kane}\ and\ \citenamefont {Mele}(2005)}]{Kane2005}%
  \BibitemOpen
  \bibfield  {author} {\bibinfo {author} {\bibfnamefont {C.~L.}\ \bibnamefont
  {Kane}}\ and\ \bibinfo {author} {\bibfnamefont {E.~J.}\ \bibnamefont
  {Mele}},\ }\href {\doibase 10.1103/PhysRevLett.95.226801} {\bibfield
  {journal} {\bibinfo  {journal} {Phys. Rev. Lett.}\ }\textbf {\bibinfo
  {volume} {95}},\ \bibinfo {pages} {226801} (\bibinfo {year}
  {2005})}\BibitemShut {NoStop}%
\bibitem [{\citenamefont {K{\"o}nig}\ \emph {et~al.}(2007)\citenamefont
  {K{\"o}nig}, \citenamefont {Wiedmann}, \citenamefont {Br{\"u}ne},
  \citenamefont {Roth}, \citenamefont {Buhmann}, \citenamefont {Molenkamp},
  \citenamefont {Qi},\ and\ \citenamefont {Zhang}}]{Molenkamp2007}%
  \BibitemOpen
  \bibfield  {author} {\bibinfo {author} {\bibfnamefont {M.}~\bibnamefont
  {K{\"o}nig}}, \bibinfo {author} {\bibfnamefont {S.}~\bibnamefont {Wiedmann}},
  \bibinfo {author} {\bibfnamefont {C.}~\bibnamefont {Br{\"u}ne}}, \bibinfo
  {author} {\bibfnamefont {A.}~\bibnamefont {Roth}}, \bibinfo {author}
  {\bibfnamefont {H.}~\bibnamefont {Buhmann}}, \bibinfo {author} {\bibfnamefont
  {L.~W.}\ \bibnamefont {Molenkamp}}, \bibinfo {author} {\bibfnamefont {X.-L.}\
  \bibnamefont {Qi}}, \ and\ \bibinfo {author} {\bibfnamefont {S.-C.}\
  \bibnamefont {Zhang}},\ }\href {https://doi.org/10.1126/science.1148047}
  {\bibfield  {journal} {\bibinfo  {journal} {Science}\ }\textbf {\bibinfo
  {volume} {318}},\ \bibinfo {pages} {766} (\bibinfo {year}
  {2007})}\BibitemShut {NoStop}%
\bibitem [{\citenamefont {Roth}\ \emph {et~al.}(2009)\citenamefont {Roth},
  \citenamefont {Br{\"u}ne}, \citenamefont {Buhmann}, \citenamefont
  {Molenkamp}, \citenamefont {Maciejko}, \citenamefont {Qi},\ and\
  \citenamefont {Zhang}}]{Molenkamp2009}%
  \BibitemOpen
  \bibfield  {author} {\bibinfo {author} {\bibfnamefont {A.}~\bibnamefont
  {Roth}}, \bibinfo {author} {\bibfnamefont {C.}~\bibnamefont {Br{\"u}ne}},
  \bibinfo {author} {\bibfnamefont {H.}~\bibnamefont {Buhmann}}, \bibinfo
  {author} {\bibfnamefont {L.~W.}\ \bibnamefont {Molenkamp}}, \bibinfo {author}
  {\bibfnamefont {J.}~\bibnamefont {Maciejko}}, \bibinfo {author}
  {\bibfnamefont {X.-L.}\ \bibnamefont {Qi}}, \ and\ \bibinfo {author}
  {\bibfnamefont {S.-C.}\ \bibnamefont {Zhang}},\ }\href
  {https://doi.org/10.1126/science.1174736} {\bibfield  {journal} {\bibinfo
  {journal} {Science}\ }\textbf {\bibinfo {volume} {325}},\ \bibinfo {pages}
  {294} (\bibinfo {year} {2009})}\BibitemShut {NoStop}%
\bibitem [{\citenamefont {Wiedenmann}\ \emph {et~al.}(2016)\citenamefont
  {Wiedenmann}, \citenamefont {Bocquillon}, \citenamefont {Deacon},
  \citenamefont {Hartinger}, \citenamefont {Herrmann}, \citenamefont
  {Klapwijk}, \citenamefont {Maier}, \citenamefont {Ames}, \citenamefont
  {Br{\"u}ne}, \citenamefont {Gould} \emph {et~al.}}]{wiedenmann2016}%
  \BibitemOpen
  \bibfield  {author} {\bibinfo {author} {\bibfnamefont {J.}~\bibnamefont
  {Wiedenmann}}, \bibinfo {author} {\bibfnamefont {E.}~\bibnamefont
  {Bocquillon}}, \bibinfo {author} {\bibfnamefont {R.~S.}\ \bibnamefont
  {Deacon}}, \bibinfo {author} {\bibfnamefont {S.}~\bibnamefont {Hartinger}},
  \bibinfo {author} {\bibfnamefont {O.}~\bibnamefont {Herrmann}}, \bibinfo
  {author} {\bibfnamefont {T.~M.}\ \bibnamefont {Klapwijk}}, \bibinfo {author}
  {\bibfnamefont {L.}~\bibnamefont {Maier}}, \bibinfo {author} {\bibfnamefont
  {C.}~\bibnamefont {Ames}}, \bibinfo {author} {\bibfnamefont {C.}~\bibnamefont
  {Br{\"u}ne}}, \bibinfo {author} {\bibfnamefont {C.}~\bibnamefont {Gould}},
  \emph {et~al.},\ }\href {https://www.nature.com/articles/ncomms10303}
  {\bibfield  {journal} {\bibinfo  {journal} {Nature communications}\ }\textbf
  {\bibinfo {volume} {7}},\ \bibinfo {pages} {10303} (\bibinfo {year}
  {2016})}\BibitemShut {NoStop}%
\bibitem [{\citenamefont {Jacquet}\ \emph {et~al.}(2018)\citenamefont
  {Jacquet}, \citenamefont {Rech}, \citenamefont {Jonckheere}, \citenamefont
  {Zazunov},\ and\ \citenamefont {Martin}}]{jacquet2018}%
  \BibitemOpen
  \bibfield  {author} {\bibinfo {author} {\bibfnamefont {R.}~\bibnamefont
  {Jacquet}}, \bibinfo {author} {\bibfnamefont {J.}~\bibnamefont {Rech}},
  \bibinfo {author} {\bibfnamefont {T.}~\bibnamefont {Jonckheere}}, \bibinfo
  {author} {\bibfnamefont {A.}~\bibnamefont {Zazunov}}, \ and\ \bibinfo
  {author} {\bibfnamefont {T.}~\bibnamefont {Martin}},\ }\href
  {https://arxiv.org/abs/1807.09112} {\bibfield  {journal} {\bibinfo  {journal}
  {arXiv:1807.09112}\ } (\bibinfo {year} {2018})}\BibitemShut {NoStop}%
\bibitem [{\citenamefont {Cr{\'e}pin}\ \emph {et~al.}(2015)\citenamefont
  {Cr{\'e}pin}, \citenamefont {Burset},\ and\ \citenamefont
  {Trauzettel}}]{crepin2015}%
  \BibitemOpen
  \bibfield  {author} {\bibinfo {author} {\bibfnamefont {F.}~\bibnamefont
  {Cr{\'e}pin}}, \bibinfo {author} {\bibfnamefont {P.}~\bibnamefont {Burset}},
  \ and\ \bibinfo {author} {\bibfnamefont {B.}~\bibnamefont {Trauzettel}},\
  }\href {https://journals.aps.org/prb/abstract/10.1103/PhysRevB.92.100507}
  {\bibfield  {journal} {\bibinfo  {journal} {Physical Review B}\ }\textbf
  {\bibinfo {volume} {92}},\ \bibinfo {pages} {100507} (\bibinfo {year}
  {2015})}\BibitemShut {NoStop}%
\bibitem [{\citenamefont {Fleckenstein}\ \emph {et~al.}(2018)\citenamefont
  {Fleckenstein}, \citenamefont {Ziani},\ and\ \citenamefont
  {Trauzettel}}]{fleckenstein2018}%
  \BibitemOpen
  \bibfield  {author} {\bibinfo {author} {\bibfnamefont {C.}~\bibnamefont
  {Fleckenstein}}, \bibinfo {author} {\bibfnamefont {N.~T.}\ \bibnamefont
  {Ziani}}, \ and\ \bibinfo {author} {\bibfnamefont {B.}~\bibnamefont
  {Trauzettel}},\ }\href
  {https://journals.aps.org/prb/abstract/10.1103/PhysRevB.97.134523} {\bibfield
   {journal} {\bibinfo  {journal} {Phys. Rev. B}\ }\textbf {\bibinfo {volume}
  {97}},\ \bibinfo {pages} {134523} (\bibinfo {year} {2018})}\BibitemShut
  {NoStop}%
\bibitem [{\citenamefont {Xiao}\ \emph {et~al.}(2016)\citenamefont {Xiao},
  \citenamefont {Liu}, \citenamefont {Liu}, \citenamefont {Ai}, \citenamefont
  {Yang},\ and\ \citenamefont {Zhou}}]{xiao2016}%
  \BibitemOpen
  \bibfield  {author} {\bibinfo {author} {\bibfnamefont {X.}~\bibnamefont
  {Xiao}}, \bibinfo {author} {\bibfnamefont {Y.}~\bibnamefont {Liu}}, \bibinfo
  {author} {\bibfnamefont {Z.}~\bibnamefont {Liu}}, \bibinfo {author}
  {\bibfnamefont {G.}~\bibnamefont {Ai}}, \bibinfo {author} {\bibfnamefont
  {S.~A.}\ \bibnamefont {Yang}}, \ and\ \bibinfo {author} {\bibfnamefont
  {G.}~\bibnamefont {Zhou}},\ }\href
  {https://aip.scitation.org/doi/abs/10.1063/1.4940239} {\bibfield  {journal}
  {\bibinfo  {journal} {Appl. Phys. Lett.}\ }\textbf {\bibinfo {volume}
  {108}},\ \bibinfo {pages} {032403} (\bibinfo {year} {2016})}\BibitemShut
  {NoStop}%
\bibitem [{\citenamefont {Tkachov}\ \emph {et~al.}(2015)\citenamefont
  {Tkachov}, \citenamefont {Burset}, \citenamefont {Trauzettel},\ and\
  \citenamefont {Hankiewicz}}]{Tkachov2015}%
  \BibitemOpen
  \bibfield  {author} {\bibinfo {author} {\bibfnamefont {G.}~\bibnamefont
  {Tkachov}}, \bibinfo {author} {\bibfnamefont {P.}~\bibnamefont {Burset}},
  \bibinfo {author} {\bibfnamefont {B.}~\bibnamefont {Trauzettel}}, \ and\
  \bibinfo {author} {\bibfnamefont {E.}~\bibnamefont {Hankiewicz}},\ }\href
  {\doibase 10.1103/PhysRevB.92.045408} {\bibfield  {journal} {\bibinfo
  {journal} {Phys. Rev. B}\ }\textbf {\bibinfo {volume} {92}},\ \bibinfo
  {pages} {045408} (\bibinfo {year} {2015})}\BibitemShut {NoStop}%
\bibitem [{\citenamefont {Sothmann}\ \emph {et~al.}(2017)\citenamefont
  {Sothmann}, \citenamefont {Giazotto},\ and\ \citenamefont
  {Hankiewicz}}]{sothmann2017}%
  \BibitemOpen
  \bibfield  {author} {\bibinfo {author} {\bibfnamefont {B.}~\bibnamefont
  {Sothmann}}, \bibinfo {author} {\bibfnamefont {F.}~\bibnamefont {Giazotto}},
  \ and\ \bibinfo {author} {\bibfnamefont {E.~M.}\ \bibnamefont {Hankiewicz}},\
  }\href {http://iopscience.iop.org/article/10.1088/1367-2630/aa60d4}
  {\bibfield  {journal} {\bibinfo  {journal} {New J. Phys.}\ }\textbf {\bibinfo
  {volume} {19}},\ \bibinfo {pages} {023056} (\bibinfo {year}
  {2017})}\BibitemShut {NoStop}%
\bibitem [{\citenamefont {Qi}\ \emph {et~al.}(2008)\citenamefont {Qi},
  \citenamefont {Hughes},\ and\ \citenamefont {Zhang}}]{Qi2008}%
  \BibitemOpen
  \bibfield  {author} {\bibinfo {author} {\bibfnamefont {X.-L.}\ \bibnamefont
  {Qi}}, \bibinfo {author} {\bibfnamefont {T.~L.}\ \bibnamefont {Hughes}}, \
  and\ \bibinfo {author} {\bibfnamefont {S.-C.}\ \bibnamefont {Zhang}},\
  }\href@noop {} {\bibfield  {journal} {\bibinfo  {journal} {Nature Phys.}\
  }\textbf {\bibinfo {volume} {4}},\ \bibinfo {pages} {273} (\bibinfo {year}
  {2008})}\BibitemShut {NoStop}%
\bibitem [{\citenamefont {Maciejko}\ \emph {et~al.}(2010)\citenamefont
  {Maciejko}, \citenamefont {Qi},\ and\ \citenamefont {Zhang}}]{Maciejko2010}%
  \BibitemOpen
  \bibfield  {author} {\bibinfo {author} {\bibfnamefont {J.}~\bibnamefont
  {Maciejko}}, \bibinfo {author} {\bibfnamefont {X.-L.}\ \bibnamefont {Qi}}, \
  and\ \bibinfo {author} {\bibfnamefont {S.-C.}\ \bibnamefont {Zhang}},\ }\href
  {\doibase 10.1103/PhysRevB.82.155310} {\bibfield  {journal} {\bibinfo
  {journal} {Phys. Rev. B}\ }\textbf {\bibinfo {volume} {82}},\ \bibinfo
  {pages} {155310} (\bibinfo {year} {2010})}\BibitemShut {NoStop}%
\bibitem [{\citenamefont {B\"uttiker}(1986)}]{Buttiker1986}%
  \BibitemOpen
  \bibfield  {author} {\bibinfo {author} {\bibfnamefont {M.}~\bibnamefont
  {B\"uttiker}},\ }\href {\doibase 10.1103/PhysRevLett.57.1761} {\bibfield
  {journal} {\bibinfo  {journal} {Phys. Rev. Lett.}\ }\textbf {\bibinfo
  {volume} {57}},\ \bibinfo {pages} {1761} (\bibinfo {year}
  {1986})}\BibitemShut {NoStop}%
\bibitem [{\citenamefont {B\"uttiker}(1992)}]{Buttiker1992}%
  \BibitemOpen
  \bibfield  {author} {\bibinfo {author} {\bibfnamefont {M.}~\bibnamefont
  {B\"uttiker}},\ }\href {\doibase 10.1103/PhysRevB.46.12485} {\bibfield
  {journal} {\bibinfo  {journal} {Phys. Rev. B}\ }\textbf {\bibinfo {volume}
  {46}},\ \bibinfo {pages} {12485} (\bibinfo {year} {1992})}\BibitemShut
  {NoStop}%
\bibitem [{\citenamefont {Datta}(1997)}]{Datta1997}%
  \BibitemOpen
  \bibfield  {author} {\bibinfo {author} {\bibfnamefont {S.}~\bibnamefont
  {Datta}},\ }\href@noop {} {\emph {\bibinfo {title} {Electronic transport in
  mesoscopic systems}}}\ (\bibinfo  {publisher} {Cambridge university press},\
  \bibinfo {year} {1997})\BibitemShut {NoStop}%
\bibitem [{\citenamefont {Lambert}\ and\ \citenamefont
  {Raimondi}(1998)}]{Lambert1998}%
  \BibitemOpen
  \bibfield  {author} {\bibinfo {author} {\bibfnamefont {C.~J.}\ \bibnamefont
  {Lambert}}\ and\ \bibinfo {author} {\bibfnamefont {R.}~\bibnamefont
  {Raimondi}},\ }\href {http://stacks.iop.org/0953-8984/10/i=5/a=003}
  {\bibfield  {journal} {\bibinfo  {journal} {J. Phys. Condens. Matter}\
  }\textbf {\bibinfo {volume} {10}},\ \bibinfo {pages} {901} (\bibinfo {year}
  {1998})}\BibitemShut {NoStop}%
\bibitem [{\citenamefont {Beenakker}(1991)}]{beenakker1991}%
  \BibitemOpen
  \bibfield  {author} {\bibinfo {author} {\bibfnamefont {C.~W.~J.}\
  \bibnamefont {Beenakker}},\ }\href {\doibase 10.1103/PhysRevLett.67.3836}
  {\bibfield  {journal} {\bibinfo  {journal} {Phys. Rev. Lett.}\ }\textbf
  {\bibinfo {volume} {67}},\ \bibinfo {pages} {3836} (\bibinfo {year}
  {1991})}\BibitemShut {NoStop}%
\bibitem [{\citenamefont {Beenakker}(1992)}]{beenakker1992}%
  \BibitemOpen
  \bibfield  {author} {\bibinfo {author} {\bibfnamefont {C.}~\bibnamefont
  {Beenakker}},\ }in\ \href@noop {} {\emph {\bibinfo {booktitle} {Transport
  Phenomena in Mesoscopic Systems}}}\ (\bibinfo  {publisher} {Springer},\
  \bibinfo {year} {1992})\ pp.\ \bibinfo {pages} {235--253}\BibitemShut
  {NoStop}%
\bibitem [{Note1()}]{Note1}%
  \BibitemOpen
  \bibinfo {note} {This can be done by adding a floating metal pad over those
  edge modes which will induce electron decoherence.}\BibitemShut {Stop}%
\bibitem [{Note2()}]{Note2}%
  \BibitemOpen
  \bibinfo {note} {The angular notation for the action of the local fields
  imply a $2\pi $-periodicity of their actions.}\BibitemShut {Stop}%
\bibitem [{\citenamefont {Beenakker}\ \emph {et~al.}(2003)\citenamefont
  {Beenakker}, \citenamefont {Emary}, \citenamefont {Kindermann},\ and\
  \citenamefont {van Velsen}}]{Beenakker2003}%
  \BibitemOpen
  \bibfield  {author} {\bibinfo {author} {\bibfnamefont {C.~W.~J.}\
  \bibnamefont {Beenakker}}, \bibinfo {author} {\bibfnamefont {C.}~\bibnamefont
  {Emary}}, \bibinfo {author} {\bibfnamefont {M.}~\bibnamefont {Kindermann}}, \
  and\ \bibinfo {author} {\bibfnamefont {J.~L.}\ \bibnamefont {van Velsen}},\
  }\href {\doibase 10.1103/PhysRevLett.91.147901} {\bibfield  {journal}
  {\bibinfo  {journal} {Phys. Rev. Lett.}\ }\textbf {\bibinfo {volume} {91}},\
  \bibinfo {pages} {147901} (\bibinfo {year} {2003})}\BibitemShut {NoStop}%
\bibitem [{\citenamefont {Dolcini}\ \emph {et~al.}(2015)\citenamefont
  {Dolcini}, \citenamefont {Houzet},\ and\ \citenamefont
  {Meyer}}]{Dolcini2015}%
  \BibitemOpen
  \bibfield  {author} {\bibinfo {author} {\bibfnamefont {F.}~\bibnamefont
  {Dolcini}}, \bibinfo {author} {\bibfnamefont {M.}~\bibnamefont {Houzet}}, \
  and\ \bibinfo {author} {\bibfnamefont {J.~S.}\ \bibnamefont {Meyer}},\ }\href
  {\doibase 10.1103/PhysRevB.92.035428} {\bibfield  {journal} {\bibinfo
  {journal} {Phys. Rev. B}\ }\textbf {\bibinfo {volume} {92}},\ \bibinfo
  {pages} {035428} (\bibinfo {year} {2015})}\BibitemShut {NoStop}%
\bibitem [{\citenamefont {Campagnano}\ \emph {et~al.}(2015)\citenamefont
  {Campagnano}, \citenamefont {Lucignano}, \citenamefont {Giuliano},\ and\
  \citenamefont {Tagliacozzo}}]{campagnano2015}%
  \BibitemOpen
  \bibfield  {author} {\bibinfo {author} {\bibfnamefont {G.}~\bibnamefont
  {Campagnano}}, \bibinfo {author} {\bibfnamefont {P.}~\bibnamefont
  {Lucignano}}, \bibinfo {author} {\bibfnamefont {D.}~\bibnamefont {Giuliano}},
  \ and\ \bibinfo {author} {\bibfnamefont {A.}~\bibnamefont {Tagliacozzo}},\
  }\href
  {http://iopscience.iop.org/article/10.1088/0953-8984/27/20/205301/meta}
  {\bibfield  {journal} {\bibinfo  {journal} {J. Phys. Condens. Matter}\
  }\textbf {\bibinfo {volume} {27}},\ \bibinfo {pages} {205301} (\bibinfo
  {year} {2015})}\BibitemShut {NoStop}%
\bibitem [{\citenamefont {Minutillo}\ \emph {et~al.}(2018)\citenamefont
  {Minutillo}, \citenamefont {Giuliano}, \citenamefont {Lucignano},
  \citenamefont {Tagliacozzo},\ and\ \citenamefont
  {Campagnano}}]{minutillo2018}%
  \BibitemOpen
  \bibfield  {author} {\bibinfo {author} {\bibfnamefont {M.}~\bibnamefont
  {Minutillo}}, \bibinfo {author} {\bibfnamefont {D.}~\bibnamefont {Giuliano}},
  \bibinfo {author} {\bibfnamefont {P.}~\bibnamefont {Lucignano}}, \bibinfo
  {author} {\bibfnamefont {A.}~\bibnamefont {Tagliacozzo}}, \ and\ \bibinfo
  {author} {\bibfnamefont {G.}~\bibnamefont {Campagnano}},\ }\href
  {https://journals.aps.org/prb/abstract/10.1103/PhysRevB.98.144510} {\bibfield
   {journal} {\bibinfo  {journal} {Physical Review B}\ }\textbf {\bibinfo
  {volume} {98}},\ \bibinfo {pages} {144510} (\bibinfo {year}
  {2018})}\BibitemShut {NoStop}%
\bibitem [{Note3()}]{Note3}%
  \BibitemOpen
  \bibinfo {note} {We tested that numerically for $\eta \to 0$ one obtain again
  the analytical result.}\BibitemShut {Stop}%
\bibitem [{\citenamefont {Yokoyama}\ \emph {et~al.}(2013)\citenamefont
  {Yokoyama}, \citenamefont {Eto},\ and\ \citenamefont
  {V.~Nazarov}}]{Yokoyama2013}%
  \BibitemOpen
  \bibfield  {author} {\bibinfo {author} {\bibfnamefont {T.}~\bibnamefont
  {Yokoyama}}, \bibinfo {author} {\bibfnamefont {M.}~\bibnamefont {Eto}}, \
  and\ \bibinfo {author} {\bibfnamefont {Y.}~\bibnamefont {V.~Nazarov}},\
  }\href {https://doi.org/10.7566/JPSJ.82.054703} {\bibfield  {journal}
  {\bibinfo  {journal} {J. Phys. Soc. Jpn.}\ }\textbf {\bibinfo {volume}
  {82}},\ \bibinfo {pages} {054703} (\bibinfo {year} {2013})}\BibitemShut
  {NoStop}%
\bibitem [{\citenamefont {Yokoyama}\ \emph {et~al.}(2014)\citenamefont
  {Yokoyama}, \citenamefont {Eto},\ and\ \citenamefont {Nazarov}}]{Eto2014}%
  \BibitemOpen
  \bibfield  {author} {\bibinfo {author} {\bibfnamefont {T.}~\bibnamefont
  {Yokoyama}}, \bibinfo {author} {\bibfnamefont {M.}~\bibnamefont {Eto}}, \
  and\ \bibinfo {author} {\bibfnamefont {Y.~V.}\ \bibnamefont {Nazarov}},\
  }\href {https://link.aps.org/doi/10.1103/PhysRevB.89.195407} {\bibfield
  {journal} {\bibinfo  {journal} {Phys. Rev. B}\ }\textbf {\bibinfo {volume}
  {89}},\ \bibinfo {pages} {195407} (\bibinfo {year} {2014})}\BibitemShut
  {NoStop}%
\bibitem [{\citenamefont {Marra}\ \emph {et~al.}(2016)\citenamefont {Marra},
  \citenamefont {Citro},\ and\ \citenamefont {Braggio}}]{Marra2016}%
  \BibitemOpen
  \bibfield  {author} {\bibinfo {author} {\bibfnamefont {P.}~\bibnamefont
  {Marra}}, \bibinfo {author} {\bibfnamefont {R.}~\bibnamefont {Citro}}, \ and\
  \bibinfo {author} {\bibfnamefont {A.}~\bibnamefont {Braggio}},\ }\href
  {\doibase 10.1103/PhysRevB.93.220507} {\bibfield  {journal} {\bibinfo
  {journal} {Phys. Rev. B}\ }\textbf {\bibinfo {volume} {93}},\ \bibinfo
  {pages} {220507} (\bibinfo {year} {2016})}\BibitemShut {NoStop}%
\bibitem [{\citenamefont {Nava}\ \emph {et~al.}(2016)\citenamefont {Nava},
  \citenamefont {Giuliano}, \citenamefont {Campagnano},\ and\ \citenamefont
  {Giuliano}}]{Nava2016}%
  \BibitemOpen
  \bibfield  {author} {\bibinfo {author} {\bibfnamefont {A.}~\bibnamefont
  {Nava}}, \bibinfo {author} {\bibfnamefont {R.}~\bibnamefont {Giuliano}},
  \bibinfo {author} {\bibfnamefont {G.}~\bibnamefont {Campagnano}}, \ and\
  \bibinfo {author} {\bibfnamefont {D.}~\bibnamefont {Giuliano}},\ }\href
  {\doibase 10.1103/PhysRevB.94.205125} {\bibfield  {journal} {\bibinfo
  {journal} {Phys. Rev. B}\ }\textbf {\bibinfo {volume} {94}},\ \bibinfo
  {pages} {205125} (\bibinfo {year} {2016})}\BibitemShut {NoStop}%
\bibitem [{\citenamefont {Mellars}\ and\ \citenamefont
  {B\'eri}(2016)}]{Mellars2016}%
  \BibitemOpen
  \bibfield  {author} {\bibinfo {author} {\bibfnamefont {E.}~\bibnamefont
  {Mellars}}\ and\ \bibinfo {author} {\bibfnamefont {B.}~\bibnamefont
  {B\'eri}},\ }\href {\doibase 10.1103/PhysRevB.94.174508} {\bibfield
  {journal} {\bibinfo  {journal} {Phys. Rev. B}\ }\textbf {\bibinfo {volume}
  {94}},\ \bibinfo {pages} {174508} (\bibinfo {year} {2016})}\BibitemShut
  {NoStop}%
\bibitem [{\citenamefont {Nesterov}\ \emph {et~al.}(2016)\citenamefont
  {Nesterov}, \citenamefont {Houzet},\ and\ \citenamefont
  {Meyer}}]{Nesterov2016}%
  \BibitemOpen
  \bibfield  {author} {\bibinfo {author} {\bibfnamefont {K.~N.}\ \bibnamefont
  {Nesterov}}, \bibinfo {author} {\bibfnamefont {M.}~\bibnamefont {Houzet}}, \
  and\ \bibinfo {author} {\bibfnamefont {J.~S.}\ \bibnamefont {Meyer}},\ }\href
  {\doibase 10.1103/PhysRevB.93.174502} {\bibfield  {journal} {\bibinfo
  {journal} {Phys. Rev. B}\ }\textbf {\bibinfo {volume} {93}},\ \bibinfo
  {pages} {174502} (\bibinfo {year} {2016})}\BibitemShut {NoStop}%
\bibitem [{Note4()}]{Note4}%
  \BibitemOpen
  \bibinfo {note} {The lowest order $(1-\eta )^4$ accounts for the single shot
  CP process, where the CP is splitted at one barrier, taking an $(1-\eta )^2$
  factor, and another factor when it recombines on the other
  barrier}\BibitemShut {NoStop}%
\bibitem [{Sch()}]{Schwabl}%
  \BibitemOpen
  \href@noop {} {\emph {\bibinfo {title} {Schwabl, Franz. Advanced quantum
  mechanics. Springer Science Business Media, 2005.}}}\BibitemShut {Stop}%
\end{thebibliography}%

%

%
%
%
%
%
%
%
%
%
%

\end{document}